\documentstyle[preprint,prabib,aps]{revtex}

\begin{document}
\draft
\date{\today}
\title{Black Tori Solutions in Einstein and 5D Gravity }
\author{Sergiu I. Vacaru \thanks{
E-Mails : sergiu$_{-}$vacaru@yahoo.com,\ sergiuvacaru@venus.nipne.ro}}
\address{Physics Department, CSU Fresno,\ Fresno, CA 93740-8031, USA, \\
and \\
Centro Multidisciplinar de Astrofisica - CENTRA, Departamento de Fisica,\\
Instituto Superior Tenico, Av. Rovisco Pais 1, Lisboa, 1049-001, Portugal}

\maketitle

\begin{abstract}
The 'anholonomic frame' method \cite{v,v2,vth} is applied for
constructing new classes of exact solutions of vacuum Einstein
equations with off--diagonal metrics in 4D and 5D gravity. We
examine several black tori solutions generated by anholonomic
transforms with non--trivial topology of the Schwarzshild metric,
which have a static toroidal horizon. We define ansatz and
parametrizations which contain warping factors, running constants
(in time and extra dimension coordinates) and effective nonlinear
gravitational polarizations. Such anisotropic vacuum toroidal
metrics, the first example was given in \cite{v}, differ
substantially from the well known toroidal black holes
\cite{lemos} which were constructed as non--vacuum solutions of
the Einstein--Maxwell gravity with cosmological constant.
Finally, we analyze two anisotropic 5D and 4D black tori solutions
with cosmological constant.

\vskip5pt

Pacs 04.50.+h, 11.25.M, 11.10.Kk, 12.10.-g
\end{abstract}

\section{Introduction}

Black hole - torus systems \cite{putten} and toroidal black holes
\cite{lemos,v} became objects of astrophysical interest since it
was shown that they are inevitable outcome of complete
gravitational collapse of a massive star, cluster of stars, or
can be present in the center of galactic systems.

Black hole and black tori solutions appear naturally as exact
solutions in general relativity and extra dimension gravity
theories. Such solutions can be constructed in both
assymptotically flat spacetiems and in spacetimes with
cosmological constant,  posses a specific supersymmetry and could
be with toroidal, cylindrical or planar topology \cite{lemos}.

String theory suggests that we may live in a fundamentally higher
dimensional spacetime \cite{stringb}. The recent approaches  are
based on the assumption that our Universe is realized as a three
dimensional (in brief, 3D) brane, modeling a 4D
pseudo--Riemannian spacetime, embedded in the 5D anti--de Sitter
($AdS_{5}$) bulk spacetime. It was proposed  in the Rundall and
Sundrum (RS) papers \cite{rs} that  such models could be with
relatively large extra dimension as a way to solve the hierarchy
problem in high energy physics.

In the present paper we explore possible black tori solutions in
5D and 4D gravity. We obtain a new class of exact solutions to
the 5D vacuum Einstein equations in the bulk, which have toroidal
horizons and are related via anholonomic transforms with toroidal
deformations of the Schwarzshild solutions. The solutions could
be with warped factors, running constants and anisotropic
gravitational polarizations. We than consider 4D black tori
solutions and generalize both 5D and 4D constructions for
spacetimes with cosmological constant.

We also discuss implications of existence of such anisotropic
 black tori solutions with non-trivial topology to the extra
 dimension gravity and general relativity theory. We prove that
 warped metrics can be obtained from vacuum 5D gravity and
 not only from a brane configurations with specific
 energy--momentum tensor.

We apply the  Salam, Strathee and Peracci \cite{sal} idea on a
gauge field like status of the coefficients of off--diagonal
metrics in extra dimension gravitity and develop it in a new
fashion by applying the method of anholonomic frames with
associated nonlinear connections  on 5D and 4D (pseudo) Riemannian
spaces \cite{v,v2,vth}.

We use the term 'locally anisotropic' spacetime (or 'anisotropic' spacetime)
for a 5D (4D) pseudo-Riemannian spacetime provided with an anholonomic frame
structure with mixed holonomic and anholonomic variables. The anisotropy of
gravitational interactions is modeled by off--diagonal metrics, or,
equivalently, by theirs diagonalized analogs given with respect to
anholonomic frames.

The paper is organized as follow:\ In Sec. II we consider two
off--diagonal metric ansatz, construct the corresponding exact
solutions of 5D vacuum Einstein equations and illustrate the
possibility of extension by introducing matter fields and
 the cosmological constant term.
  In Sec. III we construct two classes of 5D
anisotropic black tori solutions and consider  subclasses and
reparemetizations of such solutions in order to generate new ones.
Sec. IV is devoted to 4D black tori solutions. In Sec. V we extend
the approach for anisotropic 5D and 4D spacetimes with
cosmological constant  and give two examples of 5D and 4D
anisotropic black tori solution. Finally, in Sec. VI, we conclude
 and discuss the obtained results.

\section{Off--Diagonal Metric Ansatz}

We introduce the basic denotations and two ansatz for off--diagonal 5D
metrics (see details in Refs. \cite{v,v2,vth}) to be applied in definition
of anisotropic black tori solutions.

Let us consider a 5D pseudo--Riemannian spacetime provided with local
coordinates $u^{\alpha }=(x^{i},y^{4}=v,y^{5}),$ for indices like $%
i,j,k,..=1,2,3$ and $a,b,...=4,5.$ The $x^{i}$--coordinates are called
holonomic and $y^{a}$--coordinates are called anholonomic (anisotropic);
they are given respectively with respect to some holonomic and anholonomic
subframes (see the formulae (\ref{dder1}) and (\ref{anh})). Every coordinate
$x^{i}$ or $y^{a}$ could be a time \ like, 3D space, or the 5th (extra
dimensional) coordinate; we shall fix on necessity different
parametrizations.

We investigate two classes of 5D metrics:

The first type of metrics are given by a line element
\begin{equation}
ds^{2}=g_{\alpha \beta }\left( x^{i},v\right) du^{\alpha }du^{\beta }
\label{metric}
\end{equation}
with the metric coefficients $g_{\alpha \beta }$ parametrized with respect
to the coordinate co--frame $du^{\alpha },$ being dual to $\partial _{\alpha
}=\partial /\partial u^{\alpha },$ $\ $by an off--diagonal matrix (ansatz)

{
\begin{equation}
\left[
\begin{array}{ccccc}
g_{1}+w_{1}^{\ 2}h_{4}+n_{1}^{\ 2}h_{5} & w_{1}w_{2}h_{4}+n_{1}n_{2}h_{5} &
w_{1}w_{3}h_{4}+n_{1}n_{3}h_{5} & w_{1}h_{4} & n_{1}h_{5} \\
w_{1}w_{2}h_{4}+n_{1}n_{2}h_{5} & g_{2}+w_{2}^{\ 2}h_{4}+n_{2}^{\ 2}h_{5} &
w_{2}w_{3}h_{4}+n_{2}n_{3}h_{5} & w_{2}h_{4} & n_{2}h_{5} \\
w_{1}w_{3}h_{4}+n_{1}n_{3}h_{5} & w_{2}w_{3}h_{4}+n_{2}n_{3}h_{5} &
g_{3}+w_{3}^{\ 2}h_{4}+n_{3}^{\ 2}h_{5} & w_{3}h_{4} & n_{3}h_{5} \\
w_{1}h_{4} & w_{2}h_{4} & w_{3}h_{4} & h_{4} & 0 \\
n_{1}h_{5} & n_{2}h_{5} & n_{3}h_{5} & 0 & h_{5}
\end{array}
\right] ,  \label{ansatz}
\end{equation}
} where the coefficients are some necessary smoothly class functions of
type:
\begin{eqnarray}
g_{1} &=&\pm 1,g_{2,3}=g_{2,3}(x^{2},x^{3}),h_{4,5}=h_{4,5}(x^{i},v),
\nonumber \\
w_{i} &=&w_{i}(x^{i},v),n_{i}=n_{i}(x^{i},v).  \nonumber
\end{eqnarray}

The second type of metrics are given by a line element (with a
conformal factor $\Omega (x^{i},v)$ and additional deformations
of the metric via coefficients $\zeta _{\hat{\imath}}(x^{i},v),$
indices with 'hat' take values like $\hat{{i}}=1,2,3,5))$ written
as
\begin{equation}
ds^{2}=\Omega ^{2}(x^{i},v)\hat{{g}}_{\alpha \beta }\left( x^{i},v\right)
du^{\alpha }du^{\beta },  \label{cmetric}
\end{equation}
were the coefficients $\hat{{g}}_{\alpha \beta }$ are parametrized by the
ansatz {\scriptsize
\begin{equation}
\left[
\begin{array}{ccccc}
g_{1}+(w_{1}^{\ 2}+\zeta _{1}^{\ 2})h_{4}+n_{1}^{\ 2}h_{5} &
(w_{1}w_{2}+\zeta _{1}\zeta _{2})h_{4}+n_{1}n_{2}h_{5} & (w_{1}w_{3}+\zeta
_{1}\zeta _{3})h_{4}+n_{1}n_{3}h_{5} & (w_{1}+\zeta _{1})h_{4} & n_{1}h_{5}
\\
(w_{1}w_{2}+\zeta _{1}\zeta _{2})h_{4}+n_{1}n_{2}h_{5} & g_{2}+(w_{2}^{\
2}+\zeta _{2}^{\ 2})h_{4}+n_{2}^{\ 2}h_{5} & (w_{2}w_{3}++\zeta _{2}\zeta
_{3})h_{4}+n_{2}n_{3}h_{5} & (w_{2}+\zeta _{2})h_{4} & n_{2}h_{5} \\
(w_{1}w_{3}+\zeta _{1}\zeta _{3})h_{4}+n_{1}n_{3}h_{5} & (w_{2}w_{3}++\zeta
_{2}\zeta _{3})h_{4}+n_{2}n_{3}h_{5} & g_{3}+(w_{3}^{\ 2}+\zeta _{3}^{\
2})h_{4}+n_{3}^{\ 2}h_{5} & (w_{3}+\zeta _{3})h_{4} & n_{3}h_{5} \\
(w_{1}+\zeta _{1})h_{4} & (w_{2}+\zeta _{2})h_{4} & (w_{3}+\zeta _{3})h_{4}
& h_{4} & 0 \\
n_{1}h_{5} & n_{2}h_{5} & n_{3}h_{5} & 0 & h_{5}+\zeta _{5}h_{4}
\end{array}
\right] .  \label{ansatzc}
\end{equation}
}

For trivial values $\Omega =1$ and $\zeta _{\hat{\imath}}=0,$ the  line
interval (\ref{cmetric}) transforms into (\ref{metric}).

The quadratic line element (\ref{metric}) with metric coefficients (\ref
{ansatz}) can be diagonalized,
\begin{equation}
\delta
s^{2}=[g_{1}(dx^{1})^{2}+g_{2}(dx^{2})^{2}+g_{3}(dx^{3})^{2}+h_{4}(\delta
v)^{2}+h_{5}(\delta y^{5})^{2}],  \label{dmetric}
\end{equation}
with respect to the anholonomic co--frame $\left( dx^{i},\delta v,\delta
y^{5}\right) ,$ where
\begin{equation}
\delta v=dv+w_{i}dx^{i}\mbox{ and }\delta y^{5}=dy^{5}+n_{i}dx^{i}
\label{ddif1}
\end{equation}
which is dual to the frame $\left( \delta _{i},\partial _{4},\partial
_{5}\right) ,$ where
\begin{equation}
\delta _{i}=\partial _{i}+w_{i}\partial _{4}+n_{i}\partial _{5}.
\label{dder1}
\end{equation}
The bases (\ref{ddif1}) and (\ref{dder1}) are considered to satisfy some
anholonomic relations of type
\begin{equation}
\delta _{i}\delta _{j}-\delta _{j}\delta _{i}=W_{ij}^{k}\delta _{k}
\label{anh}
\end{equation}
for some non--trivial values of anholonomy coefficients $W_{ij}^{k}.$ We
obtain a holonomic (coordinate) base if the coefficients $W_{ij}^{k}$ vanish.

The quadratic line element (\ref{cmetric}) with metric coefficients (\ref
{ansatzc}) can be also diagonalized,
\begin{equation}
\delta s^{2}=\Omega
^{2}(x^{i},v)[g_{1}(dx^{1})^{2}+g_{2}(dx^{2})^{2}+g_{3}(dx^{3})^{2}+h_{4}(%
\hat{{\delta }}v)^{2}+h_{5}(\delta y^{5})^{2}],  \label{cdmetric}
\end{equation}
but with respect to another anholonomic co--frame $\left( dx^{i},\hat{{%
\delta }}v,\delta y^{5}\right) ,$ with
\begin{equation}
\delta v=dv+(w_{i}+\zeta _{i})dx^{i}+\zeta _{5}\delta y^{5}\mbox{ and }%
\delta y^{5}=dy^{5}+n_{i}dx^{i}  \label{ddif2}
\end{equation}
which is dual to the frame $\left( \hat{{\delta }}_{i},\partial _{4},\hat{{%
\partial }}_{5}\right) ,$ where
\begin{equation}
\hat{{\delta }}_{i}=\partial _{i}-(w_{i}+\zeta _{i})\partial
_{4}+n_{i}\partial _{5},\hat{{\partial }}_{5}=\partial _{5}-\zeta
_{5}\partial _{4}.  \label{dder2}
\end{equation}

The nontrivial components of the 5D Ricci tensor, $R_{~\alpha }^{\beta },$
for the metric (\ref{dmetric}) given with respect to anholonomic frames (\ref
{ddif1}) and (\ref{dder1}) are
\begin{eqnarray}
R_{2}^{2} &=&R_{3}^{3}=-\frac{1}{2g_{2}g_{3}}[g_{3}^{\bullet \bullet }-\frac{%
g_{2}^{\bullet }g_{3}^{\bullet }}{2g_{2}}-\frac{(g_{3}^{\bullet })^{2}}{%
2g_{3}}+g_{2}^{^{\prime \prime }}-\frac{g_{2}^{^{\prime }}g_{3}^{^{\prime }}%
}{2g_{3}}-\frac{(g_{2}^{^{\prime }})^{2}}{2g_{2}}],  \label{ricci1a} \\
R_{4}^{4} &=&R_{5}^{5}=-\frac{\beta }{2h_{4}h_{5}},  \label{ricci2a} \\
R_{4i} &=&-w_{i}\frac{\beta }{2h_{5}}-\frac{\alpha _{i}}{2h_{5}},
\label{ricci3a} \\
R_{5i} &=&-\frac{h_{5}}{2h_{4}}\left[ n_{i}^{\ast \ast }+\gamma n_{i}^{\ast }%
\right]  \label{ricci4a}
\end{eqnarray}
where
\begin{equation}
\alpha _{i}=\partial _{i}{h_{5}^{\ast }}-h_{5}^{\ast }\partial _{i}\ln \sqrt{%
|h_{4}h_{5}|},\beta =h_{5}^{\ast \ast }-h_{5}^{\ast }[\ln \sqrt{|h_{4}h_{5}|}%
]^{\ast },\gamma =3h_{5}^{\ast }/2h_{5}-h_{4}^{\ast }/h_{4}.  \label{abc}
\end{equation}

For simplicity, the partial derivatives are denoted like $a^{\times
}=\partial a/\partial x^{1},a^{\bullet }=\partial a/\partial
x^{2},a^{^{\prime }}=\partial a/\partial x^{3},a^{\ast }=\partial a/\partial
v.$\bigskip

We obtain the same values of the Ricci tensor for the second ansatz (\ref
{cdmetric}) if there are satisfied the conditions
\begin{equation}
\hat{{\delta }}_{i}h_{4}=0\mbox{\ and\  }\hat{{\delta }}_{i}\Omega =0
\label{conf1}
\end{equation}
and the values $\zeta _{\hat{{i}}}=\left( \zeta _{{i}},\zeta _{{5}}=0\right)
$ are found as to be a unique solution of (\ref{conf1}); for instance, if
\begin{equation}
\Omega ^{q_{1}/q_{2}}=h_{4}~(q_{1}\mbox{ and }q_{2}\mbox{ are integers}),
\label{confq}
\end{equation}
the coefficients $\zeta _{{i}}$ must solve the equations \
\begin{equation}
\partial _{i}\Omega -(w_{i}+\zeta _{{i}})\Omega ^{\ast }=0.  \label{confeq}
\end{equation}

\bigskip The system of 5D vacuum Einstein equations, $R_{~\alpha }^{\beta
}=0,$ reduces to a system of nonlinear equations with separation of
variables,
\[
R_{2}^{2}=0,~R_{4}^{4}=0,~R_{4i}=0,R_{5i}=0,
\]
which together with (\ref{confeq}) can be solved in general form \cite{vth}:
For any given values of $g_{2}$ (or $g_{3})$, $h_{4}$ (or $h_{5})$ and $%
\Omega ,$ and stated boundary conditions we can define \ consequently the
set of metric coefficients $g_{3}$(or $g_{2})$, $h_{4}$ (or $%
h_{4}),w_{i},n_{i}$ and $\zeta _{{i}}.$

The introduced ansatz can be used also for constructing solutions of 5D and
4D Einstein equations with nontrivial energy-momentum tensor
\[
R_{\alpha \beta }-\frac{1}{2}g_{\alpha \beta }R=\kappa \Upsilon _{\alpha
\beta }.
\]

The non--trivial diagonal components of the Einstein tensor, $G_{\beta
}^{\alpha }=R_{\beta }^{\alpha }-\frac{1}{2}R\delta _{\beta }^{\alpha },$
for the metric (\ref{dmetric}), given with respect to anholonomic frames,
are
\begin{equation}
G_{1}^{1}=-\left( R_{2}^{2}+S_{4}^{4}\right)
,G_{2}^{2}=G_{3}^{3}=-S_{4}^{4},G_{4}^{4}=G_{5}^{5}=-R_{2}^{2}.
\label{einstdiag}
\end{equation}
So, we can extend the system of 5D vacuum Einstein equations  by introducing
matter fields for which the energy--momentum tensor $\Upsilon _{\alpha \beta
}$ given with respect to anholonomic frames satisfy the conditions
\begin{equation}
\Upsilon _{1}^{1}=\Upsilon _{2}^{2}+\Upsilon _{4}^{4},\Upsilon
_{2}^{2}=\Upsilon _{3}^{3},\Upsilon _{4}^{4}=\Upsilon _{5}^{5}.
\label{emcond}
\end{equation}

We note that, in general, the tensor $\Upsilon _{\alpha \beta }$
may be not symmetric because with respect to anholonomic frames
there are imposed constraints which makes non symmetric the Ricci
and Einstein tensors \ \cite{v,v2,vth}.

In the simplest case we can consider a ''vacuum'' source induced by a
non--vanishing 4D cosmological constant, $\Lambda .$ In order to satisfy the
conditions (\ref{emcond}) the source induced by $\Lambda $ should be in the
form $\kappa \Upsilon _{\alpha \beta }=(2\Lambda g_{11},\Lambda g_{%
\underline{\alpha }\underline{\beta }}),$ where underlined indices $%
\underline{\alpha },\underline{\beta },...$ run 4D values $2,3,4,5.$ We note
that in 4D anholonomic gravity the source $\kappa \Upsilon _{\underline{\alpha }%
\underline{\beta }}=\Lambda g_{\underline{\alpha
}\underline{\beta }}$ satisfies the equalities $\Upsilon
_{2}^{2}=\Upsilon _{3}^{3}=\Upsilon _{4}^{4}=\Upsilon _{5}^{5}.$

By straightforward computations we obtain that the nontrivial components of
the 5D Einstein equations with anisotropic cosmological constant, $%
R_{11}=2\Lambda g_{11}$ and $R_{\underline{\alpha }\underline{\beta }%
}=\Lambda g_{\underline{\alpha }\underline{\beta }},$ for the
ansatz (\ref {ansatzc}) and anholonomic metric (\ref{cdmetric})
given with respect to anholonomic frames (\ref{ddif2}) and
(\ref{dder2}), are written in a form with separated  variables:
\begin{eqnarray}
g_{3}^{\bullet \bullet }-\frac{g_{2}^{\bullet }g_{3}^{\bullet }}{2g_{2}}-%
\frac{(g_{3}^{\bullet })^{2}}{2g_{3}}+g_{2}^{^{\prime \prime }}-\frac{%
g_{2}^{^{\prime }}g_{3}^{^{\prime }}}{2g_{3}}-\frac{(g_{2}^{^{\prime }})^{2}%
}{2g_{2}} &=&2\Lambda g_{2}g_{3},  \label{ein1} \\
h_{5}^{\ast \ast }-h_{5}^{\ast }[\ln \sqrt{|h_{4}h_{5}|}]^{\ast }
&=&2\Lambda h_{4}h_{5},  \label{ein2} \\
w_{i}\beta +\alpha _{i} &=&0,  \label{ein3} \\
n_{i}^{\ast \ast }+\gamma n_{i}^{\ast } &=&0,  \label{ein4} \\
\partial _{i}\Omega -(w_{i}+\zeta _{{i}})\Omega ^{\ast } &=&0.  \label{einc}
\end{eqnarray}
where
\begin{equation}
\alpha _{i}=\partial _{i}{h_{5}^{\ast }}-h_{5}^{\ast }\partial _{i}\ln \sqrt{%
|h_{4}h_{5}|},\beta =2\Lambda h_{4}h_{5},\gamma =3h_{5}^{\ast
}/2h_{5}-h_{4}^{\ast }/h_{4}.  \label{abcc}
\end{equation}
In the vacuum case (with $\Lambda =0)$ these equations are compatible if $%
\beta =\alpha _{i}=0$ which results that $w_{i}\left(
x^{i},v\right) $ could be arbitrary functions; this reflects a
fredom in definition of the
holonomic coordinates. For simplicity, for vacuum solutions we shall put $%
w_{i}=0.$ Finally, we remark that we can ''select'' 4D Einstein soltutions
from an ansatz (\ref{ansatz}) or (\ref{ansatzc}) by considering that the
metric coefficients do not depend on variable $x^{1},$ which mean that in
the system of equations (\ref{ein1})--(\ref{einc}) we have to deal with 4D
values $w_{\underline{i}}\left( x^{\underline{k}},v\right) ,n_{\underline{i}%
}\left( x^{\underline{k}},v\right) ,\zeta _{{i}}\left( x^{\underline{k}%
},v\right) ,$ and $h_{4}\left( x^{\underline{k}},v\right) ,h_{5}\left( x^{%
\underline{k}},v\right) ,\Omega \left( x^{\underline{k}},v\right) .$

\section{5D Black Tori}

Our goal is to apply the anholonomic frame method as to construct such exact
solutions of vacuum (and with cosmological constant) 5D Einstein equations
as they have a static toroidal horizon for a metric ansatz (\ref{ansatz}) or
(\ref{ansatzc}) which can be diagonalized with respect to some well defined
anholonomic frames. Such solutions are defined as some anholonomic
transforms of the Schwarzshild solution to a toroidal configuration with
non--trivial topology. \ In general form, they could be defined with warped
factors, running constants (in time and extra dimension coordinate) and
nonlinear polarizations.

\subsection{Toroidal deformations of the Schwarzschild metric}

Let us consider the system of {\it \ isotropic spherical coordinates} $(\rho
,\theta ,\varphi ),$ \thinspace where the isotropic radial coordinate $\rho $
is related with the usual radial coordinate $r$ via the relation $r=\rho
\left( 1+r_{g}/4\rho \right) ^{2}$ for $r_{g}=2G_{[4]}m_{0}/c^{2}$ being the
4D gravitational radius of a point particle of mass $m_{0},$ $%
G_{[4]}=1/M_{P[4]}^{2}$ is the 4D Newton constant expressed via Plank mass $%
M_{P[4]}$ (following modern string/brane theories, $M_{P[4]}$ can be
considered as a value induced from extra dimensions). We put the light speed
constant $c=1.$ This system of coordinates is considered for the so--called
isotropic representation of the Schwarzschild solution \cite{ll}
\begin{equation}
ds^{2}=\left( \frac{\widehat{\rho }-1}{\widehat{\rho }+1}\right)
^{2}dt^{2}-\rho _{g}^{2}\left( \frac{\widehat{\rho }+1}{\widehat{\rho }}%
\right) ^{4}\left( d\widehat{\rho }^{2}+\widehat{\rho }^{2}d\theta ^{2}+%
\widehat{\rho }^{2}\sin ^{2}\theta d\varphi ^{2}\right) ,  \label{schw}
\end{equation}
where, for our further considerations, we re--scaled the isotropic radial
coordinate as $\widehat{\rho }=\rho /\rho _{g},$ with $\rho _{g}=r_{g}/4.$
The metric (\ref{schw}) is a vacuum static solution of 4D Einstein equations
with spherical symmetry describing the gravitational field of a point
particle of mass $m_{0}.$ It has a singularity for $r=0$ and a spherical
horizon for $r=r_{g},$ or, in re--scaled isotropic coordinates, for $%
\widehat{\rho }=1.$ We emphasize that this solution is parametrized by a
diagonal metric given with respect to holonomic coordinate frames.

We also introduce the {\it \ toroidal coordinates} (in our case considered
as alternatives to the isotropic radial coordinates) \cite{korn} $(\sigma
,\tau ,\varphi ),$ running values $-\pi \leq \sigma <\pi ,0\leq \tau \leq
\infty ,0\leq \varphi <2\pi ,$ which are related with the isotropic 3D
Cartezian coordinates via transforms
\begin{equation}
\tilde{x}=\frac{\widetilde{\rho }\sinh \tau }{\cosh \tau -\cos \sigma }\cos
\varphi ,\tilde{y}=\frac{\widetilde{\rho }\sinh \tau }{\cosh \tau -\cos
\sigma }\sin \varphi ,\tilde{z}=\frac{\widetilde{\rho }\sinh \sigma }{\cosh
\tau -\cos \sigma }  \label{rec}
\end{equation}
and define a toroidal hypersurface
\[
\left( \sqrt{\tilde{x}^{2}+\tilde{y}^{2}}-\widetilde{\rho }\frac{\cosh \tau
}{\sinh \tau }\right) ^{2}+\tilde{z}^{2}=\frac{\widetilde{\rho }^{2}}{\sinh
^{2}\tau }.
\]
The 3D metric on a such toroidal hypersurface is
\[
ds_{(3D)}^{2}=g_{\sigma \sigma }d\sigma ^{2}+g_{\tau \tau }d\tau
^{2}+g_{\varphi \varphi }d\varphi ^{2},
\]
where
\[
g_{\sigma \sigma }=g_{\tau \tau }=\frac{\widetilde{\rho }^{2}}{\left( \cosh
\tau -\cos \sigma \right) ^{2}},g_{\varphi \varphi }=\frac{\widetilde{\rho }%
^{2}\sinh ^{2}\tau }{\left( \cosh \tau -\cos \sigma \right) ^{2}}.
\]

We can relate the toroidal coordinates $\left( \sigma ,\tau ,\varphi \right)
$ from (\ref{rec}) with the isotropic radial coordinates $\left( \widehat{%
\rho },\theta ,\varphi \right) $, scaled by the constant $\rho _{g},$ from (%
\ref{schw}) as
\[
\widetilde{\rho }=1,\sinh ^{-1}\tau =\widehat{\rho }
\]
and transform the Schwarzschild solution into a new metric with
toroidal coordinates \ by changing the 3D radial line element
into the toroidal one and stating the $tt$--coefficient of the
metric to have a toroidal horizon. The resulting metric is
\begin{equation}
ds_{(S)}^{2}=\left( \frac{\sinh \tau -1}{\sinh \tau +1}\right)
^{2}dt^{2}-\rho _{g}^{2}\frac{\left( \sinh \tau +1\right) ^{4}}{\left( \cosh
\tau -\cos \sigma \right) ^{2}}\left( d\sigma ^{2}+d\tau ^{2}+\sinh ^{2}\tau
d\varphi ^{2}\right) ],  \label{schtor}
\end{equation}
Such deformed Schwarzchild like toroidal metric is not an exact
solution of the vacuum Einstein equations, but at long radial
distances it transform into usual Schwarzchild solution with the
3D line element parametrized by toroidal coordinates.

For our further considerations we introduce two Classes (A and B)
of 4D auxiliary pseudo-Riemannian metrics, also given in toroidal
coordinates, being some conformal transforms of (\ref{schtor}),
like
\[
ds_{(S)}^{2}=\Omega _{A,B}\left( \sigma ,\tau \right) ds_{(A,B)}^{2}
\]
but which are not supposed to be solutions of the Einstein equations:

\begin{itemize}
\item  Metric of Class A:
\begin{equation}
ds_{(A)}^{2}=-d\sigma ^{2}-d\tau ^{2}+a(\tau )d\varphi ^{2}+b(\sigma ,\tau
)dt^{2}],  \label{auxm1}
\end{equation}
where
\[
a(\tau )=-\sinh ^{2}\tau \mbox{ and }b(\sigma ,\tau )=-\frac{\left( \sinh
\tau -1\right) ^{2}\left( \cosh \tau -\cos \sigma \right) ^{2}}{\rho
_{g}^{2}\left( \sinh \tau +1\right) ^{6}},
\]
which results in the metric (\ref{schtor}) by multiplication on the
conformal factor
\begin{equation}
\Omega _{A}\left( \sigma ,\tau \right) =\rho _{g}^{2}\frac{\left( \sinh \tau
+1\right) ^{4}}{\left( \cosh \tau -\cos \sigma \right) ^{2}}.  \label{auxm1c}
\end{equation}

\item  Metric of Class B:
\begin{equation}
ds_{(B)}^{2}=g(\tau )\left( d\sigma ^{2}+d\tau ^{2}\right) -d\varphi
^{2}+f(\sigma ,\tau )dt^{2},  \label{auxm2}
\end{equation}
where
\[
g(\tau )=-\sinh ^{-2}\tau \mbox{ and }f(\sigma ,\tau )=\rho _{g}^{2}\left(
\frac{\sinh ^{2}\tau -1}{\cosh \tau -\cos \sigma }\right) ^{2},
\]
which results in the metric (\ref{schtor}) by multiplication on the
conformal factor
\[
\Omega _{B}\left( \sigma ,\tau \right) =\rho _{g}^{-2}\frac{\left( \cosh
\tau -\cos \sigma \right) ^{2}}{\left( \sinh \tau +1\right) ^{2}}.
\]
\end{itemize}

We shall use the metrics (\ref{schtor}), (\ref{auxm1}) and
(\ref{auxm2}) in order to  generate exact solutions of the
Einstein equations with toroidal horizons and anisotropic
polarizations and running of constants by performing
corresponding anholonomic transforms as the solutions will have a
horizon parametrized by a torus hypersurface  and gravitational
(extra dimensional, or nonlinear 4D)
renormalizations  of the constant $\rho _{g}$ of the Schwarzschild solution, $%
\rho _{g}\rightarrow \overline{\rho }_{g}=\omega \rho _{g},$
where the dependence of the function $\omega $ on some holonomic
or anholonomic coordinates will depend on the type of anisotropy.
For some solutions we shall treat $\omega $ as a factor modeling
running of the gravitational constant,
induced, induced from extra dimension, in another cases we will consider $%
\omega $ as a nonlinear gravitational polarization which models
some anisotropic distributions of masses and matter fields and/or
anholonomic vacuum gravitational interactions.

\subsection{Toroidal 5D metrics of Class A}

In this subsection we consider four classes of 5D vacuum
solutions which are related to the metric of Class A
(\ref{auxm1}) and to the toroidally deformed  Schwarzshild metric
 (\ref{schtor}).

Let us parametrize the 5D coordinates as $\left( x^{1}=\chi ,x^{2}=\sigma
,x^{3}=\tau ,y^{4}=v,y^{5}=p\right) ,$ where the solutions with the
so--called $\varphi $--anisotropy will be constructed for $\left( v=\varphi
,p=t\right) $ and the solutions with $t$--anisotropy will be stated for $%
\left( v=t,p=\varphi \right) $ (in brief, we write \ respectively, $\varphi $%
--solutions and $t$--solutions).

\subsubsection{Class A of vacuum solutions with ansatz (\ref{ansatz}):}

We take an off--diagonal metric ansatz of type (\ref{ansatz}) (equivalently,
(\ref{metric})) by reprezenting
\[
g_{1}=\pm 1,g_{2}=-1,g_{3}=-1,h_{4}=\eta _{4}(\sigma ,\tau
,v)h_{4(0)}(\sigma ,\tau )\mbox{
and }h_{5}=\eta _{5}(\sigma ,\tau ,v)h_{5(0)}(\sigma ,\tau ),
\]
where $\eta _{4,5}(\sigma ,\tau ,v)$ are corresponding ''gravitational
renormalizations'' of the metric coefficients $h_{4,5(0)}(\sigma ,\tau ).$
For $\varphi $--solutions we state $h_{4(0)}=a(\tau )$ and $%
h_{5(0)}=b(\sigma ,\tau )$ (inversely, for \ $t$--solutions, $%
h_{4(0)}=b(\sigma ,\tau )$ and $h_{5(0)}=a(\sigma ,\tau )).$ \

Next we consider a renormalized gravitational 'constant' $\overline{\rho }%
_{g}=\omega \rho _{g},$ were for $\varphi $--solutions\ the receptivity $%
\omega =\omega \left( \sigma ,\tau ,v\right) $ is included in the
graviational polarization $\eta _{5}$ as $\eta _{5}=\left[ \omega \left(
\sigma ,\tau ,\varphi \right) \right] ^{-2},$ or for $t$--solutions is
included in $\eta _{4},$ when $\eta _{4}=\left[ \omega \left( \sigma ,\tau
,t\right) \right] ^{-2}.$ We can construct an exact solution of the 5D
vacuum Einstein equations if, for explicit dependencies on anisotropic
coordinate, the metric coefficients $h_{4}$ and $h_{5}$ are related by
 the equation (\ref{ein2}), which in its turn imposes a corresponding relation
between $\eta _{4}$ and $\eta _{5},$%
\[
\eta _{4}h_{4(0)}=h_{(0)}^{2}h_{5(0)}\left[ \left( \sqrt{|\eta _{5}|}\right)
^{\ast }\right] ^{2},~h_{(0)}^{2}=const.
\]
In result, we express the polarizations $\eta _{4}$ and $\eta _{5}$ via the
value of receptivity $\omega ,$
\begin{equation}
\eta _{4}\left( \chi ,\sigma ,\tau ,\varphi \right) =h_{(0)}^{2}\frac{%
b(\sigma ,\tau )}{a(\tau )}\left\{ \left[ \omega ^{-1}\left( \chi ,\sigma
,\tau ,\varphi \right) \right] ^{\ast }\right\} ^{2},\eta _{5}\left( \chi
,\sigma ,\tau ,\varphi \right) =\omega ^{-2}\left( \chi ,\sigma ,\tau
,\varphi \right) ,  \label{etap}
\end{equation}
for $\varphi $--solutions , and
\begin{equation}
\eta _{4}\left( \chi ,\sigma ,\tau ,t\right) =\omega ^{-2}\left( \chi
,\sigma ,\tau ,t\right) ,\eta _{5}\left( \chi ,\sigma ,\tau ,t\right)
=h_{(0)}^{-2}\frac{b(\sigma ,\tau )}{a(\tau )}\left[ \int dt\omega
^{-1}\left( \chi ,\sigma ,\tau ,t\right) \right] ^{2},  \label{etat}
\end{equation}
for $t$--solutions, where $a(\tau )$ and $b(\sigma ,\tau )$ are those from (%
\ref{auxm1}).

For vacuum configurations, following (\ref{ein3}), we put $w_{i}=0.$ The
next step is to find the values of $n_{i}$ by introducing $h_{4}=\eta
_{4}h_{4(0)}$ and $h_{5}=\eta _{5}h_{5(0)}$ into the formula \ (\ref{ein4}),
which, for convenience, is expressed via general coefficients $\eta _{4}$
and $\eta _{5}.$ After two integrations on variable $v,$ we obtain the exact
solution
\begin{eqnarray}
n_{k}(\sigma ,\tau ,v) &=&n_{k[1]}\left( \sigma ,\tau \right)
+n_{k[2]}\left( \sigma ,\tau \right) \int [\eta _{4}/(\sqrt{|\eta _{5}|}%
)^{3}]dv,~\eta _{5}^{\ast }\neq 0;  \label{nel} \\
&=&n_{k[1]}\left( \sigma ,\tau \right) +n_{k[2]}\left( \sigma ,\tau \right)
\int \eta _{4}dv,\qquad ~\eta _{5}^{\ast }=0;  \nonumber \\
&=&n_{k[1]}\left( \sigma ,\tau \right) +n_{k[2]}\left( \sigma ,\tau \right)
\int [1/(\sqrt{|\eta _{5}|})^{3}]dv,~\eta _{4}^{\ast }=0,  \nonumber
\end{eqnarray}
with the functions $n_{k[2]}\left( \sigma ,\tau \right) $ are defined as to
contain the values $h_{(0)}^{2},$ $a(\tau )$ and $b(\sigma ,\tau ).$

\bigskip By introducing the formulas (\ref{etap}) for $\varphi $--solutions
(or (\ref{etat}) for $t$--solutions) and fixing some boundary condition, in
order to state the values of coefficients $n_{k[1,2]}\left( \sigma ,\tau
\right) $ we can express the ansatz components $n_{k}\left( \sigma ,\tau
,\varphi \right) $ as integrals of some functions of $\omega \left( \sigma
,\tau ,\varphi \right) $ and $\partial _{\varphi }\omega \left( \sigma ,\tau
,\varphi \right) $ (or, we can express the ansatz components $n_{k}\left(
\sigma ,\tau ,t\right) $ as integrals of some functions of $\omega \left(
\sigma ,\tau ,t\right) $ and $\partial _{t}\omega \left( \sigma ,\tau
,t\right) ).$ We do not present an explicit form of such formulas because
they depend on the type of receptivity $\omega =\omega \left( \sigma ,\tau
,v\right) ,$ which must be defined experimentally, or from some quantum
models of gravity in the quasi classical limit. We preserved a general
dependence on coordinates $\left( \sigma ,\tau \right) $ which reflect the
fact that there is a freedom in fixing holonomic coordinates (for instance,
on a toroidal hypersurface and its extensions to 4D and 5D spacetimes). \
For simplicity, we write that $n_{i}$ are some functionals of $\{\sigma
,\tau ,\omega \left( \sigma ,\tau ,v\right) ,\omega ^{\ast }\left( \sigma
,\tau ,v\right) \}$
\[
n_{i}\{\sigma ,\tau ,\omega ,\omega ^{\ast }\}=n_{i}\{\sigma ,\tau ,\omega
\left( \sigma ,\tau ,v\right) ,\omega ^{\ast }\left( \sigma ,\tau ,v\right)
\}.
\]

In conclusion, we constructed two exact solutions of the 5D vacuum Einstein
equations, defined by the ansatz (\ref{ansatz}) with coordinates and
coefficients stated by the data:
\begin{eqnarray}
\mbox{$\varphi$--solutions} &:&(x^{1}=\chi ,x^{2}=\sigma ,x^{3}=\tau
,y^{4}=v=\varphi ,y^{5}=p=t),g_{1}=\pm 1,  \nonumber \\
g_{2} &=&-1,g_{3}=-1,h_{4(0)}=a(\tau ),h_{5(0)}=b(\sigma ,\tau ),%
\mbox{see
(\ref{auxm1})};  \nonumber \\
h_{4} &=&\eta _{4}(\sigma ,\tau ,\varphi )h_{4(0)},h_{5}=\eta _{5}(\sigma
,\tau ,\varphi )h_{5(0)},  \nonumber \\
\eta _{4} &=&h_{(0)}^{2}\frac{b(\sigma ,\tau )}{a(\tau )}\left\{ \left[
\omega ^{-1}\left( \chi ,\sigma ,\tau ,\varphi \right) \right] ^{\ast
}\right\} ^{2},\eta _{5}=\omega ^{-2}\left( \chi ,\sigma ,\tau ,\varphi
\right) ,  \nonumber \\
w_{i} &=&0,n_{i}\{\sigma ,\tau ,\omega ,\omega ^{\ast }\}=n_{i}\{\sigma
,\tau ,\omega \left( \sigma ,\tau ,\varphi \right) ,\omega ^{\ast }\left(
\sigma ,\tau ,\varphi \right) \}.  \label{sol5p1}
\end{eqnarray}
and
\begin{eqnarray}
\mbox{$t$--solutions} &:&(x^{1}=\chi ,x^{2}=\sigma ,x^{3}=\tau
,y^{4}=v=t,y^{5}=p=\varphi ),g_{1}=\pm 1,  \nonumber \\
g_{2} &=&-1,g_{3}=-1,h_{4(0)}=b(\sigma ,\tau ),h_{5(0)}=a(\tau ),%
\mbox{see
(\ref{auxm1})};  \nonumber \\
h_{4} &=&\eta _{4}(\sigma ,\tau ,t)h_{4(0)},h_{5}=\eta _{5}(\sigma ,\tau
,t)h_{5(0)},  \nonumber \\
\eta _{4} &=&\omega ^{-2}\left( \chi ,\sigma ,\tau ,t\right) ,\eta
_{5}=h_{(0)}^{-2}\frac{b(\sigma ,\tau )}{a(\tau )}\left[ \int dt~\omega
^{-1}\left( \chi ,\sigma ,\tau ,t\right) \right] ^{2},  \nonumber \\
w_{i} &=&0,n_{i}\{\sigma ,\tau ,\omega ,\omega ^{\ast }\}=n_{i}\{\sigma
,\tau ,\omega \left( \sigma ,\tau ,t\right) ,\omega ^{\ast }\left( \sigma
,\tau ,t\right) \}.  \label{sol5t1}
\end{eqnarray}

Both types of solutions have a horizon parametrized by torus hypersurface
(as the condition of vanishing of \ the ''time'' metric coefficient states,
i. e. when the function $b(\sigma ,\tau )=0)$. $\ $These solutions are
generically anholonomic (anisotropic) because in the locally isotropic
limit, when $\eta _{4},\eta _{5},$ $\omega \rightarrow 1$ and $%
n_{i}\rightarrow 0,$ they reduce to the coefficients of the metric (\ref
{auxm1}). The last one is not an exact solution of 4D vacuum Einstein
equations, but it is a conformal transform of the 4D Schwarzschild metric
deformed to a torus horizon, with a further trivial extension to 5D. With
respect to the anholonomic frames adapted to the coefficients $n_{i}$ (see (%
\ref{ddif1})), the obtained solutions have diagonal metric coefficients
being very similar to the metric (\ref{schtor}) written in toroidal
coordinates. We can treat such solutions as black tori with the mass
distributed linearly on the circle which can not transformed in a point, in
the center of torus.

The solutions are constructed as to have singularities on the mentioned
circle. \ The initial data for anholonomic frames and the chosen
configuration of gravitational interactions in the bulk lead to deformed
toroidal horizons even for static configurations. The solutions admit
anisotropic polarizations on toroidal coordinates $\left( \sigma ,\tau
\right) $ and running of constants on time $t$ and/or on extra dimension
coordinate $\chi $. Such renormalizations of constants are defined by the
nonlinear configuration of the 5D vacuum gravitational field and depend on
introduced receptivity function $\omega \left( \sigma ,\tau ,v\right) $
which is to be considered an intrinsic characteristics of the 5D vacuum
gravitational 'ether', emphasizing the possibility \ of nonlinear
self--polarization of gravitational fields.

Finally, we point that the data (\ref{sol5p1}) and (\ref{sol5t1})
parametrize two very different classes of solutions. The first one is for
static 5D vacuum black tori configurations with explicit dependence on
anholonomic coordinate $\varphi $ and possible renormalizations on the rest
of 3D space coordinates $\sigma $ and $\tau $ and on the 5th coordinate $%
\chi .$ The second class of solutions is similar to the static ones but with
an emphasized anholonomic running on time of constants and with possible
anisotropic dependencies on coordinates $(\sigma ,\tau ,\chi ).$

\subsubsection{Class A of vacuum solutions with ansatz (\ref{ansatzc}):}

We construct here 5D vacuum $\varphi $-- and $t$--solutions
parametrized by an ansatz with conformal factor $\Omega (\sigma
,\tau ,v)$ (see (\ref {ansatzc}) and (\ref{cdmetric})). Let us
consider conformal factors parametrized as $\Omega =\Omega
_{\lbrack 0]}(\sigma ,\tau )\Omega _{\lbrack 1]}(\sigma ,\tau
,v).$ We can generate from the data (\ref{sol5p1}) (or (\ref
{sol5t1})) an exact solution of vacuum Einstein equations if
there are satisfied the conditions (\ref{conf1}), (\ref{confq})
and (\ref{confeq}), i. e.
\[
\Omega _{\lbrack 0]}^{q_{1}/q_{2}}\Omega _{\lbrack 1]}^{q_{1}/q_{2}}=\eta
_{4}h_{4(0)},
\]
for some integers $q_{1}$ and $q_{2},$ and there are defined the second
anisotropy coefficients
\[
\zeta _{i}=\left( \partial _{i}\ln |\Omega _{\lbrack 0]}\right) |)~\left(
\ln |\Omega _{\lbrack 1]}|\right) ^{\ast }+\left( \Omega _{\lbrack 1]}^{\ast
}\right) ^{-1}\partial _{i}\Omega _{\lbrack 1]}.
\]
So, taking a $\varphi $-- or $t$--solution with corresponding values of $%
h_{4}=\eta _{4}h_{4(0)},$\ for some $q_{1}$ and $q_{2},$ we obtain new exact
solutions, called in brief, $\varphi _{c}$-- or $t_{c}$--solutions (with the
index ''c'' pointing to an ansatz with conformal factor), of the vacuum 5D
Einstein equations given in explicit form by the data:
\begin{eqnarray}
\mbox{$\varphi_c$--solutions} &:&(x^{1}=\chi ,x^{2}=\sigma ,x^{3}=\tau
,y^{4}=v=\varphi ,y^{5}=p=t),g_{1}=\pm 1,  \nonumber \\
g_{2} &=&-1,g_{3}=-1,h_{4(0)}=a(\tau ),h_{5(0)}=b(\sigma ,\tau ),%
\mbox{see
(\ref{auxm1})};  \nonumber \\
h_{4} &=&\eta _{4}(\sigma ,\tau ,\varphi )h_{4(0)},h_{5}=\eta _{5}(\sigma
,\tau ,\varphi )h_{5(0)},  \nonumber \\
\eta _{4} &=&h_{(0)}^{2}\frac{b(\sigma ,\tau )}{a(\tau )}\left\{ \left[
\omega ^{-1}\left( \chi ,\sigma ,\tau ,\varphi \right) \right] ^{\ast
}\right\} ^{2},\eta _{5}=\omega ^{-2}\left( \chi ,\sigma ,\tau ,\varphi
\right) , \label{sol5pc} \\
w_{i} &=&0,n_{i}\{\sigma ,\tau ,\omega ,\omega ^{\ast }\}=n_{i}\{\sigma
,\tau ,\omega \left( \sigma ,\tau ,\varphi \right) ,\omega ^{\ast }\left(
\sigma ,\tau ,\varphi \right) \},\Omega =\Omega _{\lbrack 0]}(\sigma ,\tau
)\Omega _{\lbrack 1]}(\sigma ,\tau ,\varphi )   \nonumber  \\
\zeta _{i} &=&\left( \partial _{i}\ln |\Omega _{\lbrack 0]}\right) |)~\left(
\ln |\Omega _{\lbrack 1]}|\right) ^{\ast }+\left( \Omega _{\lbrack 1]}^{\ast
}\right) ^{-1}\partial _{i}\Omega _{\lbrack 1]},\eta _{4}a=\Omega _{\lbrack
0]}^{q_{1}/q_{2}}(\sigma ,\tau )\Omega _{\lbrack 1]}^{q_{1}/q_{2}}(\sigma
,\tau ,\varphi ).  \nonumber
\end{eqnarray}
and
\begin{eqnarray}
\mbox{$t_c$--solutions} &:&(x^{1}=\chi ,x^{2}=\sigma ,x^{3}=\tau
,y^{4}=v=t,y^{5}=p=\varphi ),g_{1}=\pm 1,  \nonumber \\
g_{2} &=&-1,g_{3}=-1,h_{4(0)}=b(\sigma ,\tau ),h_{5(0)}=a(\tau ),%
\mbox{see
(\ref{auxm1})};  \nonumber \\
h_{4} &=&\eta _{4}(\sigma ,\tau ,t)h_{4(0)},h_{5}=\eta _{5}(\sigma ,\tau
,t)h_{5(0)},  \nonumber \\
\eta _{4} &=&\omega ^{-2}\left( \chi ,\sigma ,\tau ,t\right) ,\eta
_{5}=h_{(0)}^{-2}\frac{b(\sigma ,\tau )}{a(\tau )}\left[ \int dt~\omega
^{-1}\left( \chi ,\sigma ,\tau ,t\right) \right] ^{2},  \label{sol5tc} \\
w_{i} &=&0,n_{i}\{\sigma ,\tau ,\omega ,\omega ^{\ast }\}=n_{i}\{\sigma
,\tau ,\omega \left( \sigma ,\tau ,t\right) ,\omega ^{\ast }\left( \sigma
,\tau ,t\right) \},\Omega =\Omega _{\lbrack 0]}(\sigma ,\tau )\Omega
_{\lbrack 1]}(\sigma ,\tau ,t) \nonumber  \\
\zeta _{i} &=&\left( \partial _{i}\ln |\Omega _{\lbrack 0]}\right) |)~\left(
\ln |\Omega _{\lbrack 1]}|\right) ^{\ast }+\left( \Omega _{\lbrack 1]}^{\ast
}\right) ^{-1}\partial _{i}\Omega _{\lbrack 1]},\eta _{4}a=\Omega _{\lbrack
0]}^{q_{1}/q_{2}}(\sigma ,\tau )\Omega _{\lbrack 1]}^{q_{1}/q_{2}}(\sigma
,\tau ,t).  \nonumber
\end{eqnarray}

These solutions have two very interesting properties: 1) they admit a warped
factor on the 5th coordinate, like $\Omega _{\lbrack 1]}^{q_{1}/q_{2}}\sim
\exp [-k|\chi |],$ which in our case is constructed for an anisotropic 5D
vacuum gravitational configuration and not following a brane configuration
like in Refs. \cite{rs}; 2) we can impose such conditions on the receptivity
$\omega \left( \sigma ,\tau ,v\right) $ as to obtain in the locally
isotropic limit just the toroidally deformed Schwarzschild metric (\ref
{schtor}) trivially embedded into the 5D spacetime.

We analyze the second property in details. We have to chose the conformal
factor as to be satisfied three conditions:
\begin{equation}
\Omega _{\lbrack 0]}^{q_{1}/q_{2}}=\Omega _{A},\Omega _{\lbrack
1]}^{q_{1}/q_{2}}\eta _{4}=1,\Omega _{\lbrack 1]}^{q_{1}/q_{2}}\eta _{5}=1,
\label{cond1a}
\end{equation}
were $\Omega _{A}$ is that from (\ref{auxm1c}). The last two conditions are
possible if
\begin{equation}
\eta _{4}^{-q_{1}/q_{2}}\eta _{5}=1,  \label{cond2}
\end{equation}
which selects a specific form of receptivity $\omega \left( x^{i},v\right) .$
\ Putting into (\ref{cond2}) the values $\eta _{4}$ and $\eta _{5}$
respectively from (\ref{sol5pc}), or (\ref{sol5tc}), we obtain some
differential, or integral, relations of the unknown $\omega \left( \sigma
,\tau ,v\right) ,$ which results that
\begin{eqnarray}
\omega \left( \sigma ,\tau ,\varphi \right) &=&\left( 1-q_{1}/q_{2}\right)
^{-1-q_{1}/q_{2}}\left[ h_{(0)}^{-1}\sqrt{|a/b|}\varphi +\omega _{\lbrack
0]}\left( \sigma ,\tau \right) \right] ,\mbox{ for }\varphi _{c}%
\mbox{--solutions};  \label{cond1} \\
\omega \left( \sigma ,\tau ,t\right) &=&\left[ \left( q_{1}/q_{2}-1\right)
h_{(0)}\sqrt{|a/b|}t+\omega _{\lbrack 1]}\left( \sigma ,\tau \right) \right]
^{1-q_{1}/q_{2}},\mbox{ for }t_{c}\mbox{--solutions},  \nonumber
\end{eqnarray}
for some arbitrary functions $\omega _{\lbrack 0]}\left( \sigma ,\tau
\right) $ and $\omega _{\lbrack 1]}\left( \sigma ,\tau \right) .$ \ So,
receptivities of particular form like (\ref{cond1}) allow us to obtain in
the locally isotropic limit just the toroidally deformed Schwarzschild
metric.

We conclude this subsection:\ the  vacuum 5D metrics solving the
Einstein equations describe a nonlinear gravitational dynamics
which under some particular boundary conditions and
parametrizations of metric's coefficients can model anisotropic,
topologically not--trivial, solutions transforming, in a
corresponding locally isotropic limit, in some toroidal or
ellipsoidal deformations of the well known exact solutions like
Schwarzschild, Reissner-N\"{o}rdstrom, Taub NUT, various type of
wormhole, solitonic and disk solutions (see details in Refs.
\cite{v,v2,vth}). We emphasize that, in general, an anisotropic
solution (parametrized by an off--diagonal ansatz) could not have
a locally isotropic limit to a diagonal metric with respect to
some holonomic coordinate frames. This was proved in exlicit form
by chosing a configuration with toroidal symmetry.

\subsection{Toroidal 5D metrics of Class B}

In this subsection we construct and analyze another two classes of 5D vacuum
solutions which are related to the metric of Class B (\ref{auxm2}) and which
can be reduced to the toroidally deformed Schwarzshild metric (\ref{schtor})
by corresponding parametrizations of receptivity $\omega \left( \sigma ,\tau
,v\right) $. We emphasize that because the function $g(\sigma ,\tau )$ from (%
\ref{auxm2}) is not a solution of equation(\ref{ein1}) we introduce an
auxiliary factor $\varpi $ $(\sigma ,\tau )$ for which $\varpi g$ becames a
such solution, then we consider conformal factors parametrized as $\Omega
=\varpi ^{-1}(\sigma ,\tau )$ $\Omega _{\lbrack 2]}\left( \sigma ,\tau
,v\right) $ and find solutions parametrized by the ansatz (\ref{ansatzc})
and anholonomic metric interval (\ref{cdmetric}).

Besause the method of definition of such solutions is similar to that from
previous subsection, in our further considerations we shall omit
computations and present directly the data which select the respective
configurations for $\varphi _{c}$--solutions and $t_{c}$--solutions.

The Class B of 5D solutions with conformal factor are parametrized by the
data:

\begin{eqnarray}
\mbox{$\varphi_c$--solutions} &:&(x^{1}=\chi ,x^{2}=\sigma ,x^{3}=\tau
,y^{4}=v=\varphi ,y^{5}=p=t),g_{1}=\pm 1,  \nonumber \\
g_{2} &=&g_{3}=\varpi (\sigma ,\tau )g(\sigma ,\tau ),h_{4(0)}=-\varpi
(\sigma ,\tau ),h_{5(0)}=\varpi (\sigma ,\tau )f(\sigma ,\tau ),%
\mbox{see
(\ref{auxm2})};  \nonumber \\
\varpi &=&g^{-1}(\sigma ,\tau )\varpi _{0}\exp [a_{2}\sigma +a_{3}\tau
],~\varpi _{0},a_{2},a_{3}=const;  \nonumber \\
h_{4} &=&\eta _{4}(\sigma ,\tau ,\varphi )h_{4(0)},h_{5}=\eta _{5}(\sigma
,\tau ,\varphi )h_{5(0)},  \nonumber \\
\eta _{4} &=&-h_{(0)}^{2}f(\sigma ,\tau )\left\{ \left[ \omega ^{-1}\left(
\chi ,\sigma ,\tau ,\varphi \right) \right] ^{\ast }\right\} ^{2},\eta
_{5}=\omega ^{-2}\left( \chi ,\sigma ,\tau ,\varphi \right) ,  \label{sol5p}
\\
w_{i} &=&0,n_{i}\{\sigma ,\tau ,\omega ,\omega ^{\ast }\}=n_{i}\{\sigma
,\tau ,\omega \left( \sigma ,\tau ,\varphi \right) ,\omega ^{\ast }\left(
\sigma ,\tau ,\varphi \right) \},\Omega =\varpi ^{-1}(\sigma ,\tau )\Omega
_{\lbrack 2]}(\sigma ,\tau ,\varphi )  \nonumber \\
\zeta _{i} &=&\partial _{i}\ln |\varpi |)~\left( \ln |\Omega _{\lbrack
2]}|\right) ^{\ast }+\left( \Omega _{\lbrack 2]}^{\ast }\right)
^{-1}\partial _{i}\Omega _{\lbrack 2]},\eta _{4}=-\varpi
^{-q_{1}/q_{2}}(\sigma ,\tau )\Omega _{\lbrack 2]}^{q_{1}/q_{2}}(\sigma
,\tau ,\varphi ).  \nonumber
\end{eqnarray}
and
\begin{eqnarray}
\mbox{$t_c$--solutions} &:&(x^{1}=\chi ,x^{2}=\sigma ,x^{3}=\tau
,y^{4}=v=t,y^{5}=p=\varphi ),g_{1}=\pm 1,  \nonumber \\
g_{2} &=&g_{3}=\varpi (\sigma ,\tau )g(\sigma ,\tau ),h_{4(0)}=\varpi
(\sigma ,\tau )f(\sigma ,\tau ),h_{5(0)}=-\varpi (\sigma ,\tau ),%
\mbox{see
(\ref{auxm2})};  \nonumber \\
\varpi &=&g^{-1}(\sigma ,\tau )\varpi _{0}\exp [a_{2}\sigma +a_{3}\tau
],~\varpi _{0},a_{2},a_{3}=const,  \nonumber \\
h_{4} &=&\eta _{4}(\sigma ,\tau ,t)h_{4(0)},h_{5}=\eta _{5}(\sigma ,\tau
,t)h_{5(0)},  \nonumber \\
\eta _{4} &=&\omega ^{-2}\left( \chi ,\sigma ,\tau ,t\right) ,\eta
_{5}=-h_{(0)}^{-2}f(\sigma ,\tau )\left[ \int dt~\omega ^{-1}\left( \chi
,\sigma ,\tau ,t\right) \right] ^{2},  \label{sol5t} \\
w_{i} &=&0,n_{i}\{\sigma ,\tau ,\omega ,\omega ^{\ast }\}=n_{i}\{\sigma
,\tau ,\omega \left( \sigma ,\tau ,t\right) ,\omega ^{\ast }\left( \sigma
,\tau ,t\right) \},\Omega =\varpi ^{-1}(\sigma ,\tau )\Omega _{\lbrack
2]}(\sigma ,\tau ,t)  \nonumber \\
\zeta _{i} &=&\partial _{i}(\ln |\varpi |)~\left( \ln |\Omega _{\lbrack
2]}|\right) ^{\ast }+\left( \Omega _{\lbrack 2]}^{\ast }\right)
^{-1}\partial _{i}\Omega _{\lbrack 2]},\eta _{4}=-\varpi
^{-q_{1}/q_{2}}(\sigma ,\tau )\Omega _{\lbrack 2]}^{q_{1}/q_{2}}(\sigma
,\tau ,t).  \nonumber
\end{eqnarray}
where the coeffiecients $n_{i}$ can be found explicitly by introducing the
corresponding values $\eta _{4}$ and $\eta _{5}$ in formula (\ref{nel}).

By a procedure similar to the solutions of Class A (see previous subsection)
we can find the conditions when the solutions (\ref{sol5p}) and (\ref{sol5t}%
) will have in the locally anisotropic limit the toroidally deformed
Schwarzshild solutions, which impose corresponding parametrizations and
dependencies on $\Omega _{\lbrack 2]}(\sigma ,\tau ,v)$ and $\omega \left(
\sigma ,\tau ,v\right) $ like (\ref{cond1a}) and (\ref{cond1}). We omit
these formulas because, in general, for aholonomic configurations and
nonlinear solutions there are not hard arguments to prefer any holonomic
limits of such off--diagonal metrics.

Finally, in this Section, we remark that for the considered classes of black
tori solutions the so--called $t$--components of metric contain
modifications of the Schwarzschild potential
\[
\Phi =-\frac{M}{M_{P[4]}^{2}r}\mbox{ into }\Phi =-\frac{M\omega \left(
\sigma ,\tau ,v\right) }{M_{P[4]}^{2}r},
\]
where $M_{P[4]}$ is the usual 4D Plank constant, and this is given with
respect to the corresponding aholonomic frame of reference. The receptivity $%
\omega \left( \sigma ,\tau ,v\right) $ could model corrections warped on
extra dimension coordinate, $\chi ,$ which for our solutions are induced by
anholonomic vacuum gravitational interactions in the bulk and not from a
brane configuration in $AdS_{5}$ spacetime. In the vacuum case $k$ is a
constant which chareacterizes the receptivity for bulk vacuum gravitational
polarizations.

\section{4D Black Tori}

For the ansatz (\ref{ansatz}), with trivial conformal factor, a
black torus solution of 4D vacuum Einstein equations was
constructed in Ref. \cite{v}. \ The goal of this Section is to
consider some alternative variants,  with trivial or nontrivial
conformal factors and for different coordinate parametrizations
and types of anisotropies. The bulk of 5D solutions from the
previous Section are reduced into corresponding 4D ones if we
eliminate the 5th coordinate $\chi $ from\ the the off--diagonal
ansatz (\ref {ansatz}) and (\ref{ansatzc}) and corresponding
formulas and
 solutions.

\subsection{Toroidal 4D vacuum metrics of Class A}

Let us parametrize the 4D coordinates as $(x^{\underline{i}},y^{a})=\left(
x^{2}=\sigma ,x^{3}=\tau ,y^{4}=v,y^{5}=p\right) ;$ for the $\varphi $%
--solutions we shall take $\left( v=\varphi ,p=t\right) $ and for the
solutions $t$--solutions will consider $\left( v=t,p=\varphi \right) $. For
simplicity, we write down the data for solutions without proofs and
computations.

\subsubsection{Class A of vacuum solutions with ansatz (\ref{ansatz}):}

The off--diagonal metric ansatz of type (\ref{ansatz}) (equivalently, (\ref
{dmetric})) \ with the data
\begin{eqnarray}
\mbox{$\varphi$--solutions} &:&(x^{2}=\sigma ,x^{3}=\tau ,y^{4}=v=\varphi
,y^{5}=p=t)  \nonumber \\
g_{2} &=&-1,g_{3}=-1,h_{4(0)}=a(\tau ),h_{5(0)}=b(\sigma ,\tau ),%
\mbox{see
(\ref{auxm1})};  \nonumber \\
h_{4} &=&\eta _{4}(\sigma ,\tau ,\varphi )h_{4(0)},h_{5}=\eta _{5}(\sigma
,\tau ,\varphi )h_{5(0)},  \nonumber \\
\eta _{4} &=&h_{(0)}^{2}\frac{b(\sigma ,\tau )}{a(\tau )}\left\{ \left[
\omega ^{-1}\left( \sigma ,\tau ,\varphi \right) \right] ^{\ast }\right\}
^{2},\eta _{5}=\omega ^{-2}\left( \sigma ,\tau ,\varphi \right) ,  \nonumber
\\
w_{\underline{i}} &=&0,n_{\underline{i}}\{\sigma ,\tau ,\omega ,\omega
^{\ast }\}=n_{\underline{i}}\{\sigma ,\tau ,\omega \left( \sigma ,\tau
,\varphi \right) ,\omega ^{\ast }\left( \sigma ,\tau ,\varphi \right) \}.
\label{sol4p1}
\end{eqnarray}
and
\begin{eqnarray}
\mbox{$t$--solutions} &:&(x^{2}=\sigma ,x^{3}=\tau
,y^{4}=v=t,y^{5}=p=\varphi )  \nonumber \\
g_{2} &=&-1,g_{3}=-1,h_{4(0)}=b(\sigma ,\tau ),h_{5(0)}=a(\tau ),%
\mbox{see
(\ref{auxm1})};  \nonumber \\
h_{4} &=&\eta _{4}(\sigma ,\tau ,t)h_{4(0)},h_{5}=\eta _{5}(\sigma ,\tau
,t)h_{5(0)},  \nonumber \\
\eta _{4} &=&\omega ^{-2}\left( \sigma ,\tau ,t\right) ,\eta
_{5}=h_{(0)}^{-2}\frac{b(\sigma ,\tau )}{a(\tau )}\left[ \int dt~\omega
^{-1}\left( \sigma ,\tau ,t\right) \right] ^{2},  \nonumber \\
w_{\underline{i}} &=&0,n_{\underline{i}}\{\sigma ,\tau ,\omega ,\omega
^{\ast }\}=n_{\underline{i}}\{\sigma ,\tau ,\omega \left( \sigma ,\tau
,t\right) ,\omega ^{\ast }\left( \sigma ,\tau ,t\right) \}.  \label{sol4t1}
\end{eqnarray}
where the $n_{\underline{i}}$ are computed
\begin{eqnarray}
n_{k}\left( \sigma ,\tau ,v\right) &=&n_{k[1]}\left( \sigma ,\tau \right)
+n_{k[2]}\left( \sigma ,\tau \right) \int [\eta _{4}/(\sqrt{|\eta _{5}|}%
)^{3}]dv,~\eta _{5}^{\ast }\neq 0;  \label{nem4} \\
&=&n_{k[1]}\left( \sigma ,\tau \right) +n_{k[2]}\left( \sigma ,\tau \right)
\int \eta _{4}dv,\qquad ~\eta _{5}^{\ast }=0;  \nonumber \\
&=&n_{k[1]}\left( \sigma ,\tau \right) +n_{k[2]}\left( \sigma ,\tau \right)
\int [1/(\sqrt{|\eta _{5}|})^{3}]dv,~\eta _{4}^{\ast }=0.  \nonumber
\end{eqnarray}
when the integration variable is taken $v=\varphi ,$ for (\ref{sol4p1}), or $%
v=t,$ for (\ref{sol4t1}). These solutions have the same toroidal symmetries
and properties stated for their 5D analogs (\ref{sol5p1}) and for (\ref
{sol5t1}) with that difference that there are not any warped factors and
extra dimension dependencies. Such solutions defined by the formulas (\ref
{sol4p1}) and (\ref{sol4t1}) do not result in a locally isotropic limit into
an exact solution having diagonal coefficients with respect to some
holonomic coordinate frames. The data introduced in this subsection are for
generic 4D vacuum solutions of the Einstein equations parametrized by
off--diagonal metrics. The renormalization of constants and metric
coefficients have a 4D nonlinear vacuum gravitational nature and reflects a
corresponding anholonomic dynamics.

\subsubsection{Class A of vacuum solutions with ansatz (\ref{ansatzc}):}

The 4D vacuum $\varphi $-- and $t$--solutions parametrized by an ansatz with
conformal factor $\Omega (\sigma ,\tau ,v)$ (see (\ref{ansatzc}) and (\ref
{cdmetric})). Let us consider conformal factors parametrized as $\Omega
=\Omega _{\lbrack 0]}(\sigma ,\tau )\Omega _{\lbrack 1]}(\sigma ,\tau ,v).$
The data are
\begin{eqnarray}
\mbox{$\varphi_c$--solutions} &:&(x^{2}=\sigma ,x^{3}=\tau ,y^{4}=v=\varphi
,y^{5}=p=t)  \nonumber \\
g_{2} &=&-1,g_{3}=-1,h_{4(0)}=a(\tau ),h_{5(0)}=b(\sigma ,\tau ),%
\mbox{see
(\ref{auxm1})};  \nonumber \\
h_{4} &=&\eta _{4}(\sigma ,\tau ,\varphi )h_{4(0)},h_{5}=\eta _{5}(\sigma
,\tau ,\varphi )h_{5(0)}, \Omega =\Omega _{\lbrack 0]}(\sigma ,\tau )\Omega _{\lbrack 1]}(\sigma ,\tau ,\varphi ),
 \nonumber \\
\eta _{4} &=&h_{(0)}^{2}\frac{b(\sigma ,\tau )}{a(\tau )}\left\{ \left[
\omega ^{-1}\left( \sigma ,\tau ,\varphi \right) \right] ^{\ast }\right\}
^{2},\eta _{5}=\omega ^{-2}\left( \sigma ,\tau ,\varphi \right) ,
\label{sol4pc} \\
w_{i} &=&0,n_{i}\{\sigma ,\tau ,\omega ,\omega ^{\ast
}\}=n_{i}\{\sigma ,\tau ,\lambda ,\omega \left( \sigma ,\tau
,\varphi \right) ,\omega ^{\ast }\left( \sigma ,\tau ,\varphi
\right) \},  \nonumber \\
\zeta _{i} &=&\left( \partial _{i}\ln |\Omega _{\lbrack 0]}\right) |)~\left(
\ln |\Omega _{\lbrack 1]}|\right) ^{\ast }+\left( \Omega _{\lbrack 1]}^{\ast
}\right) ^{-1}\partial _{i}\Omega _{\lbrack 1]},\eta _{4}a=\Omega _{\lbrack
0]}^{q_{1}/q_{2}}(\sigma ,\tau )\Omega _{\lbrack 1]}^{q_{1}/q_{2}}(\sigma
,\tau ,\varphi ).  \nonumber
\end{eqnarray}
and
\begin{eqnarray}
\mbox{$t_c$--solutions} &:&(x^{2}=\sigma ,x^{3}=\tau
,y^{4}=v=t,y^{5}=p=\varphi )  \nonumber \\
g_{2} &=&-1,g_{3}=-1,h_{4(0)}=b(\sigma ,\tau ),h_{5(0)}=a(\tau ),%
\mbox{see
(\ref{auxm1})};  \nonumber \\
h_{4} &=&\eta _{4}(\sigma ,\tau ,t)h_{4(0)},h_{5}=\eta
_{5}(\sigma ,\tau ,t)h_{5(0)}, \Omega =\Omega _{\lbrack
0]}(\sigma ,\tau )\Omega
_{\lbrack 1]}(\sigma ,\tau ,t), \nonumber \\
\eta _{4} &=&\omega ^{-2}\left( \sigma ,\tau ,t\right) ,\eta
_{5}=h_{(0)}^{-2}\frac{b(\sigma ,\tau )}{a(\tau )}\left[ \int dt~\omega
^{-1}\left( \sigma ,\tau ,t\right) \right] ^{2},  \label{sol4tc} \\
w_{i} &=&0,n_{i}\{\sigma ,\tau ,\omega ,\omega ^{\ast
}\}=n_{i}\{\sigma ,\tau ,\omega \left( \sigma ,\tau ,t\right)
,\omega ^{\ast }\left( \sigma ,\tau ,t\right) \},
 \nonumber \\
\zeta _{i} &=&\left( \partial _{i}\ln |\Omega _{\lbrack 0]}\right) |)~\left(
\ln |\Omega _{\lbrack 1]}|\right) ^{\ast }+\left( \Omega _{\lbrack 1]}^{\ast
}\right) ^{-1}\partial _{i}\Omega _{\lbrack 1]},\eta _{4}a=\Omega _{\lbrack
0]}^{q_{1}/q_{2}}(\sigma ,\tau )\Omega _{\lbrack 1]}^{q_{1}/q_{2}}(\sigma
,\tau ,t),  \nonumber
\end{eqnarray}
where the coefficients the $n_{\underline{i}}$ are given by the same
formulas (\ref{nem4}).

Contrary to the solutions (\ref{sol4p1}) and for (\ref{sol4t1}) theirs
conformal anholonomic transforms, respectively, (\ref{sol4pc}) and (\ref
{sol4tc}), can be subjected to such parametrizations of the conformal factor
and conditions on the receptivity $\omega \left( \sigma ,\tau ,v\right) $ as
to obtain in the locally isotropic limit just the toroidally deformed
Schwarzschild metric (\ref{schtor}). These conditions are stated for $\Omega
_{\lbrack 0]}^{q_{1}/q_{2}}=\Omega _{A},$ $\Omega _{\lbrack
1]}^{q_{1}/q_{2}}\eta _{4}=1,$ $\Omega _{\lbrack 1]}^{q_{1}/q_{2}}\eta
_{5}=1,$were $\Omega _{A}$ is that from (\ref{auxm1c}), which is possible if
$\eta _{4}^{-q_{1}/q_{2}}\eta _{5}=1,$which selects a specific form of the
receptivity $\omega .$ \ Putting the values $\eta _{4}$ and $\eta _{5},$
respectively, from (\ref{sol4pc}), or (\ref{sol4tc}), we obtain some
differential, or integral, relations of the unknown $\omega \left( \sigma
,\tau ,v\right) ,$ which results that
\begin{eqnarray*}
\omega \left( \sigma ,\tau ,\varphi \right) &=&\left( 1-q_{1}/q_{2}\right)
^{-1-q_{1}/q_{2}}\left[ h_{(0)}^{-1}\sqrt{|a/b|}\varphi +\omega _{\lbrack
0]}\left( \sigma ,\tau \right) \right] ,\mbox{ for }\varphi _{c}%
\mbox{--solutions}; \\
\omega \left( \sigma ,\tau ,t\right) &=&\left[ \left( q_{1}/q_{2}-1\right)
h_{(0)}\sqrt{|a/b|}t+\omega _{\lbrack 1]}\left( \sigma ,\tau \right) \right]
^{1-q_{1}/q_{2}},\mbox{ for }t_{c}\mbox{--solutions},
\end{eqnarray*}
for some arbitrary functions $\omega _{\lbrack 0]}\left( \sigma ,\tau
\right) $ and $\omega _{\lbrack 1]}\left( \sigma ,\tau \right) .$ The
obtained formulas for $\omega \left( \sigma ,\tau ,\varphi \right) $ and $%
\omega \left( \sigma ,\tau ,t\right) $ are 4D reductions of the formulas (%
\ref{cond1a}) and (\ref{cond1}).

\subsection{Toroidal 4D vacuum metrics of Class B}

We construct another two classes of 4D vacuum solutions which are related to
the metric of Class B (\ref{auxm2}) and can be reduced to the toroidally
deformed Schwarzshild metric (\ref{schtor}) by corresponding
parametrizations of receptivity $\omega \left( \sigma ,\tau ,v\right) $. The
solutions contain a 2D conformal factor $\varpi (\sigma ,\tau )$ for which $%
\varpi g$ becomes a solution of (\ref{ein1}) and a 4D conformal factor
parametrized as $\Omega =\varpi ^{-1}$ $\Omega _{\lbrack 2]}\left( \sigma
,\tau ,v\right) $ in \ order to set the constructions into the ansatz (\ref
{ansatzc}) and anholonomic metric interval (\ref{cdmetric}).

The data selecting the 4D configurations for $\varphi _{c}$--solutions and $%
t_{c}$--solutions:

\begin{eqnarray}
\mbox{$\varphi_c$--solutions} &:&(x^{2}=\sigma ,x^{3}=\tau ,y^{4}=v=\varphi
,y^{5}=p=t)  \nonumber \\
g_{2} &=&g_{3}=\varpi (\sigma ,\tau )g(\sigma ,\tau ),h_{4(0)}=-\varpi
(\sigma ,\tau ),h_{5(0)}=\varpi (\sigma ,\tau )f(\sigma ,\tau ),%
\mbox{see
(\ref{auxm2})};  \nonumber \\
\varpi &=&g^{-1}\varpi _{0}\exp [a_{2}\sigma +a_{3}\tau ],~\varpi
_{0},a_{2},a_{3}=const;  \nonumber \\
h_{4} &=&\eta _{4}(\sigma ,\tau ,\varphi )h_{4(0)},h_{5}=\eta
_{5}(\sigma ,\tau ,\varphi )h_{5(0)}, \Omega =\varpi ^{-1}(\sigma
,\tau )\Omega
_{\lbrack 2]}(\sigma ,\tau ,\varphi )  \nonumber \\
\eta _{4} &=&-h_{(0)}^{2}f(\sigma ,\tau )\left\{ \left[ \omega ^{-1}\left(
\sigma ,\tau ,\varphi \right) \right] ^{\ast }\right\} ^{2},\eta _{5}=\omega
^{-2}\left( \sigma ,\tau ,\varphi \right) ,  \label{sol4p} \\
w_{i} &=&0,n_{i}\{\sigma ,\tau ,\omega ,\omega ^{\ast }\}=n_{i}\{\sigma
,\tau ,\omega \left( \sigma ,\tau ,\varphi \right) ,\omega ^{\ast }\left(
\sigma ,\tau ,\varphi \right) \}, \nonumber \\
\zeta _{\underline{i}} &=&\partial _{\underline{i}}\ln |\varpi |)~\left( \ln
|\Omega _{\lbrack 2]}|\right) ^{\ast }+\left( \Omega _{\lbrack 2]}^{\ast
}\right) ^{-1}\partial _{\underline{i}}\Omega _{\lbrack 2]},\eta
_{4}=-\varpi ^{-q_{1}/q_{2}}(\sigma ,\tau )\Omega _{\lbrack
2]}^{q_{1}/q_{2}}(\sigma ,\tau ,\varphi ).  \nonumber
\end{eqnarray}
and
\begin{eqnarray}
\mbox{$t_c$--solutions} &:&(x^{2}=\sigma ,x^{3}=\tau
,y^{4}=v=t,y^{5}=p=\varphi )  \nonumber \\
g_{2} &=&g_{3}=\varpi (\sigma ,\tau )g(\sigma ,\tau ),h_{4(0)}=\varpi
(\sigma ,\tau )f(\sigma ,\tau ),h_{5(0)}=-\varpi (\sigma ,\tau ),%
\mbox{see
(\ref{auxm2})};  \nonumber \\
\varpi &=&g^{-1}\varpi _{0}\exp [a_{2}\sigma +a_{3}\tau ],~\varpi
_{0},a_{2},a_{3}=const, \nonumber \\
h_{4} &=&\eta _{4}(\sigma ,\tau ,t)h_{4(0)},h_{5}=\eta
_{5}(\sigma ,\tau ,t)h_{5(0)}, \Omega =\varpi ^{-1}(\sigma ,\tau
)\Omega _{\lbrack
2]}(\sigma ,\tau ,t)  \nonumber \\
\eta _{4} &=&\omega ^{-2}\left( \sigma ,\tau ,t\right) ,\eta
_{5}=-h_{(0)}^{-2}f(\sigma ,\tau )\left[ \int dt~\omega ^{-1}\left( \sigma
,\tau ,t\right) \right] ^{2},  \label{sol4t} \\
w_{i} &=&0,n_{i}\{\sigma ,\tau ,\omega ,\omega ^{\ast }\}=n_{i}\{\sigma
,\tau ,\omega \left( \sigma ,\tau ,t\right) ,\omega ^{\ast }\left( \sigma
,\tau ,t\right) \}, \nonumber \\
\zeta _{i} &=&\partial _{i}(\ln |\varpi |)~\left( \ln |\Omega _{\lbrack
2]}|\right) ^{\ast }+\left( \Omega _{\lbrack 2]}^{\ast }\right)
^{-1}\partial _{i}\Omega _{\lbrack 2]},\eta _{4}=-\varpi
^{-q_{1}/q_{2}}(\sigma ,\tau )\Omega _{\lbrack 2]}^{q_{1}/q_{2}}(\sigma
,\tau ,t).  \nonumber
\end{eqnarray}
where the coefficients $n_{i}$ can be found explicitly by introducing the
corresponding values $\eta _{4}$ and $\eta _{5}$ in formula (\ref{nel}).

For the 4D Class B solutions, some conditions can be imposed (see previous
subsection) when the solutions (\ref{sol4p}) and (\ref{sol4t}) have in the
locally anisotropic limit the toroidally deformed Schwarzshild solution,
which imposes some specific parametrizations and dependencies on $\Omega
_{\lbrack 2]}(\sigma ,\tau ,v)$ and $\omega \left( \sigma ,\tau ,v\right) $
like (\ref{cond1a}) and (\ref{cond1}). We omit these considerations because
for aholonomic configurations and nonlinear solutions there are not
arguments to prefer any holonomic limits of such off--diagonal metrics.

We conclude this Section by noting that for the constructed classes of 4D
black tori solutions the so--called $t$--component of metric contains
modifications of the Schwarzschild potential
\[
\Phi =-\frac{M}{M_{P[4]}^{2}r}\mbox{ into }\Phi =-\frac{M\omega \left(
\sigma ,\tau ,v\right) }{M_{P[4]}^{2}r},
\]
where $M_{P[4]}$ is the usual 4D Plank constant; the metric coefficients are
given with respect to the corresponding aholonomic frame of reference. In 4D
anholonomic gravity the receptivity $\omega \left( \sigma ,\tau ,v\right) $
is considered to renormalize the mass constant. Such gravitational
self-polarizations are induced by anholonomic vacuum gravitational
interactions. They should be defined experimentally or computed following a
model of quantum gravity.

\section{The Cosmological Constant and Anisotropy}

In this Section we analyze the general properties of anholonomic Einstein
equations in 5D and 4D gravity with cosmological constant and consider two
examples of 5D and 4D exact solutions.

A non--vanishing $\Lambda $ term in the system of Einstein's equations
instricuces substantial differences becouse t $\beta \neq 0$ and, in this
case, one could be $w_{i}\neq 0;$ The equations (\ref{ein1}) and (\ref{ein2}%
) are of more general nonlinearity because of presence of the $2\Lambda
g_{2}g_{3}$ and $2\Lambda h_{4}h_{5}$ terms. In this case, the solutions
with $g_{2}=const$ and $g_{3}=const$ (and $h_{4}=const$ and $h_{5}=const)$
are not admitted. This makes more sophisticate the procedure of definition
of $g_{2}$ for a stated $g_{3}$ (or inversely, of definition of $g_{3}$ for
a stated $g_{2})$ from (\ref{ein1}) [similarly of constructing $h_{4}$ for a
given $h_{5}$ from (\ref{ein2}) and inversely], nevertheless, the separation
of variables is not affected by introduction of cosmological constant and
there is a number of possibilities to generate exact solutions.

The general properties of solutions of the system (\ref{ein1})--(\ref{einc}%
), with cosmological constant $\Lambda ,$ are stated in the form:

\begin{itemize}
\item  The general solution of equation (\ref{ein1}) is to be found from the
equation
\begin{equation}
\varpi \varpi ^{\bullet \bullet }-(\varpi ^{\bullet })^{2}+\varpi \varpi
^{^{\prime \prime }}-(\varpi ^{^{\prime }})^{2}=2\Lambda \varpi ^{3}.
\label{auxr1}
\end{equation}
for a coordinate transform coordinate transforms $x^{2,3}\rightarrow
\widetilde{x}^{2,3}\left( u,\lambda \right) $ for which
\[
g_{2}(\sigma ,\tau )(d\sigma )^{2}+g_{3}(\sigma ,\tau )(d\tau
)^{2}\rightarrow \varpi \left[ (d\widetilde{x}^{2})^{2}+\epsilon (d%
\widetilde{x}^{3})^{2}\right] ,\epsilon =\pm 1
\]
and $\varpi ^{\bullet }=\partial \varpi /\partial \widetilde{x}^{2}$ and $%
\varpi ^{^{\prime }}=\partial \varpi /\partial \widetilde{x}^{3}.$

\item  The equation (\ref{ein2}) relates two functions $h_{4}\left( \sigma
,\tau ,v\right) $ and $h_{5}\left( \sigma ,\tau ,v\right) $ with $%
h_{5}^{\ast }\neq 0.$ If the function $h_{5}$ is given we can find $h_{4}$
as a solution of
\begin{equation}
h_{4}^{\ast }+\frac{2\Lambda }{\pi }(h_{4})^{2}+2\left( \frac{\pi ^{\ast }}{%
\pi }-\pi \right) h_{4}=0,  \label{auxr2c}
\end{equation}

where $\pi =h_{5}^{\ast }/2h_{5}.$

\item  The exact solutions of (\ref{ein3}) for $\beta \neq 0$ is
\begin{eqnarray}
w_{k} &=&-\alpha _{k}/\beta ,  \label{aw} \\
&=&\partial _{k}\ln [\sqrt{|h_{4}h_{5}|}/|h_{5}^{\ast }|]/\partial _{v}\ln [%
\sqrt{|h_{4}h_{5}|}/|h_{5}^{\ast }|],  \nonumber
\end{eqnarray}
for $\partial _{v}=\partial /\partial v$ and $h_{5}^{\ast }\neq 0.$

\item  The exact solution of (\ref{ein4}) is
\begin{eqnarray}
n_{k} &=&n_{k[1]}\left( \sigma ,\tau \right) +n_{k[2]}\left( \sigma ,\tau
\right) \int [h_{4}/(\sqrt{|h_{5}|})^{3}]dv,  \label{nlambda} \\
&=&n_{k[1]}\left( \sigma ,\tau \right) +n_{k[2]}\left( \sigma ,\tau \right)
\int [1/(\sqrt{|h_{5}|})^{3}]dv,~h_{4}^{\ast }=0,  \nonumber
\end{eqnarray}
for some functions $n_{k[1,2]}\left( \sigma ,\tau \right) $ stated by
boundary conditions.

\item  The exact solution of (\ref{einc}) is given by
\begin{equation}
\zeta _{i}=-w_{i}+(\Omega ^{\ast })^{-1}\partial _{i}\Omega ,\quad \Omega
^{\ast }\neq 0,  \label{aconf4}
\end{equation}
\end{itemize}

We note that by a corresponding re--parametrizations of the conformal factor
$\Omega \left( \sigma ,\tau ,v\right) $ we can reduce (\ref{auxr1}) to
\begin{equation}
\varpi \varpi ^{\bullet \bullet }-(\varpi ^{\bullet })^{2}=2\Lambda \varpi
^{3}  \label{redaux}
\end{equation}
which gives and exact solution $\varpi =\varpi \left( \widetilde{x}%
^{2}\right) $ found from
\[
(\varpi ^{\bullet })^{2}=\varpi ^{3}\left( C\varpi ^{-1}+4\Lambda \right)
,C=const,
\]
(or, inversely, to reduce to
\[
\varpi \varpi ^{^{\prime \prime }}-(\varpi ^{^{\prime }})^{2}=2\Lambda
\varpi ^{3}
\]
with exact solution $\varpi =\varpi \left( \widetilde{x}^{3}\right) $ found
from
\[
(\varpi ^{\prime })^{2}=\varpi ^{3}\left( C\varpi ^{-1}+4\Lambda \right)
,C=const).
\]
The inverse problem of definition of $h_{5}$ for a given $h_{4}$ can be
solved in explicit form when $h_{4}^{\ast }=0,$ $h_{4}=h_{4(0)}(\sigma ,\tau
).$ In this case we have to solve
\begin{equation}
h_{5}^{\ast \ast }+\frac{(h_{5}^{\ast })^{2}}{2h_{5}}-2\Lambda
h_{4(0)}h_{5}=0,  \label{auxr2ccp}
\end{equation}
which admits exact solutions by reduction to a Bernulli equation.

The outlined properties of solutions with cosmological constant hold also
for 4D anholonomic spacetimes with ''isotropic'' cosmological constant $%
\Lambda .$ To transfer general soltuions from 5D to 4D we have to
eliminate dependencies on the coordinate $x^{1}$ and to consider
the 4D ansatz without $g_{11}$ term.

\subsection{A 5D anisotropic black torus solution with cosmological constant}

We give an example of generalization of ansiotropic black hole solutions of
Class A , constructed in the Section III as they will contain the
cosmological constant $\Lambda ;$ we extend the solutions given by the data (%
\ref{sol5pc}).

Our new 5D $\varphi $-- solution is parametrized by an ansatz with conformal
factor $\Omega (x^{i},v)$ (see (\ref{ansatzc}) and (\ref{cdmetric})) as $%
\Omega =\varpi ^{-1}(\sigma )\Omega _{\lbrack 0]}(\sigma ,\tau
)\Omega _{\lbrack 1]}(\sigma ,\tau ,v).$ The factor $\varpi
(\sigma ,\tau )$ is chosen as to be a solution of (\ref{redaux}).
This conformal data must satisfy the condition (\ref{confq}), i.
e.
\[
\varpi ^{-q_{1}/q_{2}}\Omega _{\lbrack 0]}^{q_{1}/q_{2}}\Omega _{\lbrack
1]}^{q_{1}/q_{2}}=\eta _{4}\varpi h_{4(0)}
\]
for some integers $q_{1}$ and $q_{2},$ where $\eta _{4}$ is found as $%
h_{4}=\eta _{4}\varpi h_{4(0)}$ satisfies the equation (\ref{auxr2c}) and $%
\Omega _{\lbrack 0]}(\sigma ,\tau )$ could be chosen as to obtain
in the locally isotropic limit and $\Lambda \rightarrow 0$ the
toroidally deformed Scwarzshild metric (\ref{schtor}). Choosing
$h_{5}=\eta _{5}\varpi h_{5(0)},$ $\eta _{5}h_{5(0)}$ is for the
ansatz for (\ref{sol5pc}), for which we compute the value $\pi
=h_{5}^{\ast }/2h_{5},$ we obtain from (\ref
{auxr2c}) an equation for $\eta _{4},$%
\[
\eta _{4}^{\ast }+\frac{2\Lambda }{\pi }\varpi h_{4(0)}(\eta
_{4})^{2}+2\left( \frac{\pi ^{\ast }}{\pi }-\pi \right) \eta _{4}=0
\]
which is a Bernulli equation \cite{kamke} and admit an exact solution, in
general, in non explicit form, $\eta _{4}=\eta _{4}^{[bern]}(\sigma ,\tau
,v,\Lambda ,\varpi ,\omega ,a,b),$ were we emphasize the functional
dependencies on functions $\varpi ,\omega ,a,b$ and cosmological constant $%
\Lambda .$ Having defined $\eta _{4[bern]},$ $\eta _{5}$ and
$\varpi ,$ we can compute the $\alpha _{i}$--$,\beta -,$ and
$\gamma $--coefficients, expressed as $\alpha _{i}=\alpha
_{i}^{[bern]}(\sigma ,\tau ,v,\Lambda ,\varpi ,\omega ,a,b),\beta
=\beta ^{\lbrack bern]}(\sigma ,\tau ,v,\Lambda ,\varpi ,\omega
,a,b)$ and $\gamma =\gamma ^{\lbrack bern]}(\sigma ,\tau
,v,\Lambda ,\varpi ,\omega ,a,b),$ following the formulas
(\ref{abc}).

The next step is to find
\[
w_{i}=w_{i}^{[bern]}(\sigma ,\tau ,v,\Lambda ,\varpi ,\omega ,a,b)%
\mbox{ and
}n_{i}=n_{i}^{[bern]}(\sigma ,\tau ,v,\Lambda ,\varpi ,\omega ,a,b)
\]
as for the general solutions (\ref{aw}) and (\ref{nlambda}).

At the final step we are able to compute the the second anisotropy
coefficients
\[
\zeta _{i}=-w_{i}^{[bern]}+(\partial _{i}\ln |\varpi ^{-1}\Omega _{\lbrack
0]}|)~\left( \ln |\Omega _{\lbrack 1]}|\right) ^{\ast }+\left( \Omega
_{\lbrack 1]}^{\ast }\right) ^{-1}\partial _{i}\Omega _{\lbrack 1]},
\]
which depends on an arbitrary function $\Omega _{\lbrack 0]}(\sigma ,\tau ).$
If we state $\Omega _{\lbrack 0]}(\sigma ,\tau )=\Omega _{A},$ as for $%
\Omega _{A}$ from (\ref{auxm2}), see similar details with respect to
formulas (\ref{cond1a}), (\ref{cond2}) and (\ref{cond1}).

The data for the exact solutions with cosmological constant for $v=\varphi $
can be stated in the form
\begin{eqnarray}
\mbox{$\varphi_c$--solutions} &:&(x^{1}=\chi ,x^{2}=\sigma ,x^{3}=\tau
,y^{4}=v=\varphi ,y^{5}=p=t),g_{1}=\pm 1,  \nonumber \\
g_{2} &=&\varpi (\sigma ),g_{3}=\varpi (\sigma ),h_{4(0)}=a(\tau
),h_{5(0)}=b(\sigma ,\tau ),\mbox{see (\ref{auxm1}) and  (\ref{redaux})};
\nonumber \\
h_{4} &=&\eta _{4}(\sigma ,\tau ,\varphi )\varpi (\sigma
)h_{4(0)},h_{5}=\eta _{5}(\sigma ,\tau ,\varphi )\varpi (\sigma )h_{5(0)},
\nonumber \\
\eta _{4} &=&\eta _{4}^{[bern]}(\sigma ,\tau ,v,\Lambda ,\varpi ,\omega
,a,b),\eta _{5}=\omega ^{-2}\left( \chi ,\sigma ,\tau ,\varphi \right) ,
\label{slambdap1} \\
w_{i} &=&w_{i}^{[bern]}(\sigma ,\tau ,v,\Lambda ,\varpi ,\omega
,a,b),n_{i}\{\sigma ,\tau ,\omega ,\omega ^{\ast }\}=n_{i}^{[bern]}(\sigma
,\tau ,v,\Lambda ,\varpi ,\omega ,a,b),  \nonumber \\
\Omega &=&\varpi ^{-1}(\sigma )\Omega _{\lbrack 0]}(\sigma ,\tau )\Omega
_{\lbrack 1]}(\sigma ,\tau ,\varphi ),\eta _{4}a=\Omega _{\lbrack
0]}^{q_{1}/q_{2}}(\sigma ,\tau )\Omega _{\lbrack 1]}^{q_{1}/q_{2}}(\sigma
,\tau ,\varphi ).  \nonumber \\
\zeta _{i} &=&-w_{i}^{[bern]}+\left( \partial _{i}\ln |\varpi ^{-1}\Omega
_{\lbrack 0]}\right) |)~\left( \ln |\Omega _{\lbrack 1]}|\right) ^{\ast
}+\left( \Omega _{\lbrack 1]}^{\ast }\right) ^{-1}\partial _{i}\Omega
_{\lbrack 1]}.  \nonumber
\end{eqnarray}

We note that a solution with $v=t$ can be constructed as to generalize (\ref
{sol5tc}) to the presence of $\Lambda .$ We can not present such data in
explicit form because in this case we have to define $\eta _{5}$ by solving
a solution like (\ref{ein2}) for $h_{5},$ for a given $h_{4},$ which can not
be integrated in explicit form.

The solution (\ref{slambdap1}) preserves the two  interesting
properties of (\ref{sol5pc}): 1) it admits a warped factor on the 5th coordinate, like $%
\Omega _{\lbrack 1]}^{q_{1}/q_{2}}\sim \exp [-k|\chi |],$ which
in this case is constructed for an anisotropic 5D vacuum
gravitational configuration with anisotropic cosmological
constant but not following a brane configuration like in Refs.
\cite{rs}; 2) we can impose such conditions on the receptivity
$\omega \left( \sigma ,\tau ,\varphi \right) $ as to obtain in
the locally isotropic limit just the toroidally defformed
Schwarzschild metric (\ref{schtor}) trivially embedded into the
5D spacetime.

\subsection{A 4D anisotropic black torus solution with cosmological constant}

The symplest way to construct a such solution is to take the data (\ref
{slambdap1}), for $v=\varphi $, to elliminate the variable $\chi $ and to
reduce the 5D indices to 4D ones. We obtain the 4D data:

\begin{eqnarray}
\mbox{$\varphi_c$--solutions} &:&(x^{2}=\sigma ,x^{3}=\tau ,y^{4}=v=\varphi
,y^{5}=p=t),  \nonumber \\
g_{2} &=&\varpi (\sigma ),g_{3}=\varpi (\sigma ),h_{4(0)}=a(\tau
),h_{5(0)}=b(\sigma ,\tau ),\mbox{see (\ref{auxm1}) and  (\ref{redaux})};
\nonumber \\
h_{4} &=&\eta _{4}(\sigma ,\tau ,\varphi )\varpi (\sigma
)h_{4(0)},h_{5}=\eta _{5}(\sigma ,\tau ,\varphi )\varpi (\sigma )h_{5(0)},
\nonumber \\
\eta _{4} &=&\eta _{4}^{[bern]}(\sigma ,\tau ,v,\Lambda ,\varpi ,\omega
,a,b),\eta _{5}=\omega ^{-2}\left( \sigma ,\tau ,\varphi \right) ,
\label{sot4tpanis} \\
w_{i} &=&w_{i}^{[bern]}(\sigma ,\tau ,v,\Lambda ,\varpi ,\omega
,a,b),n_{i}\{\sigma ,\tau ,\omega ,\omega ^{\ast }\}=n_{i}^{[bern]}(\sigma
,\tau ,v,\Lambda ,\varpi ,\omega ,a,b),  \nonumber \\
\Omega &=&\varpi ^{-1}(\sigma )\Omega _{\lbrack 0]}(\sigma ,\tau )\Omega
_{\lbrack 1]}(\sigma ,\tau ,\varphi ),\eta _{4}a=\Omega _{\lbrack
0]}^{q_{1}/q_{2}}(\sigma ,\tau )\Omega _{\lbrack 1]}^{q_{1}/q_{2}}(\sigma
,\tau ,\varphi ).  \nonumber \\
\zeta _{i} &=&-w_{i}^{[bern]}+\left( \partial _{i}\ln |\varpi ^{-1}\Omega
_{\lbrack 0]}\right) |)~\left( \ln |\Omega _{\lbrack 1]}|\right) ^{\ast
}+\left( \Omega _{\lbrack 1]}^{\ast }\right) ^{-1}\partial _{i}\Omega
_{\lbrack 1]}.  \nonumber
\end{eqnarray}
The solution (\ref{sot4tpanis}) describes a static black torus
solution in 4D gravity with cosmological constant, $\Lambda.$ The
parameters of solutions depends on the $\Lambda$ as well are
renormalized by nonlinear anholonomic gravitational interactions.
We can consider that the mass associtated to such toroidal
configuration can be anisotropically distributed in the interior
of the torus and gravitationally polarized.

Finally, we note that in a similar manner like in the Sections III and IV we
can construct another classes of anisotropic black holes solutions in 5D and
4D spacetimes with cosmological constants, being of Class A or Class B, with
anisotropic $\varphi $--coordinate, or anisotropic $t$--coordinate. We omit
the explicit data which are some nonlinear anholonomic generalizations of
those solutions.

\section{Conclusions and Discussion}

We have shown that static black tori solutions can be constructed
both in vacuum Einstein and five dimensional (5D) gravity. The
solutions are parametrized by  off--diagonal metric ansatz which
can diagonalized with respect to corresponding anholonomic frames
with mixtures of holonomic and anholonomic variables. Such
metrics contain a toroidal horizon being some deformations with
non-trivial topology of the Schwarzshild black hole solution.

The solutions were constructed by using the anholonomic frame
method \cite{v,v2,vth} which results in a very substantial
simplification of the Einstein equations which admit general
integrals for solutions.

The constructed black tori metrics depend on classes of two
dimensional and three dimensional functions which reflect the
freedom in definition of toroidal coordinates as well the
possibility to state by boundary conditions various
configurations with running constants, anisotropic gravitational
polarizations and (in presence of extra dimensions) with warping
geometries. The new toroidal solutions can be extended for
 spacetimes with cosmological constant.

In view of existence of such solutions, the old problem of the
status of frames in gravity theories rises once again, now in the
context of "effective" diagonalization of off--diagonal metrics
by using anholonomic transforms.  The bulk of solutions with
spherical, cylindrical and plane symmetries were constructed in
gravitational theories of diverse dimensions by using diagonal
metrics (sometimes with off--diagonal terms) given with respect
to "pure" coordinate frames. Such solutions can be equivalently
re--defined with respect to arbitrary frames of reference and
usually the  problem of fixing some reference bases in order to
state the boundary conditions is an important physical problem but
not a dynamical one. This problem becomes more sophisticate when
we deal with generic off--diagonal metrics and anholonomic frames.
In this case some 'dynamical, componets of metrics can be
transformed into "non--dynamical" components of frame bases,
which, following  a more rigorous
 mathematical approach, reflects a constrained nonlinear
 dynamics  for gravitational and matter fields with  both holonomic
 (unconstrained)   and anholonomic (constrained)  variables. In result there are
  more possibilities in definition of classes of exact solutions
  with non--trivial topology, anisotropies and nonlinear
  interactions.

The solutions obtained in this paper contain as particular cases
(for corresponding parametrizations of considered ansatz) the
'black ring' metrics with event horizon of topology $S^1 \times
S^2$  analyzed in Refs. \cite{emp}. In our case we emphasized the
presence of off--diagonal terms which results in warping,
anisotropy and running of constants. Here it should be noted that
 the generic nonlinear character of the Einstein equations written
  with respect to
  anholonomic frames connected with diagonalization of off--diagonal
   metrics allow us
 to construct different classes of exact 5D and 4D solutions with the same
 or different topology; such  solutions can define very different
  vacuum gravitational and graviational--matter field
  configurations.

The method and results presented in this paper  provide a
prescripiton on anholonomic  transforming of some known locally
isotropic solutions from a gravity/string theory into
corresponding classes of anisoropic solutions of the same, or of
an extended theory:

{\em A vacuum, or non-vacuum, solution, and metrics conformally
equivalent to a known solution, parametrized by a diagonal matrix
given with respect to a holonomic (coordinate) base, contained as
a trivial form of ansatz (\ref {ansatz}), or (\ref{ansatzc}), can
 be transformed into a metric with non-trivial topological horizons and
 then generalized to be an anisotropic solution with similar but
anisotropically renormalized physical constants and diagonal
metric coefficients, given with respect to adapted anholonomic
frames; the new anholonomic metric defines an exact solution of a
simplified form of \ the Einstein equations
(\ref{ein1})--(\ref{einc}) and (\ref {confeq}); such types of
solutions are parametrized by off--diagonal metrics if they are
re--defined with respect to usual coordinate frames}.

We emphasize that the anholonomic frame method and constructed
black tori solutions conclude in a general formalism of generating
exact solutions with off--diagonal metrics in  gravity theories
and  may have a number of applications in modern astrophysics and
string/M--theory gravity.

\subsection*{Acknowledgements}

The author thanks J.P.S. Lemos and D. Singleton for support and
collaboration. The work is supported both by a 2000--2001
California State University Legislative Award and a NATO/Portugal
fellowship grant at the Technical University of Lisbon.

\end{document}